\documentclass[11pt]{article}

\usepackage{epsfig}
\usepackage{amsmath,amssymb}
\usepackage{latexsym}
\usepackage{graphicx}
\usepackage{float}
\usepackage[english]{babel}
\usepackage[colorlinks,citecolor=blue,linkcolor=blue]{hyperref}
\usepackage{cite}
\usepackage{pdfpages}
\usepackage{slashed}
\usepackage{subcaption}
\usepackage{xcolor}
\usepackage{tikz}
\usetikzlibrary{decorations.markings}

\makeatletter
\renewcommand\section{\@startsection {section}{1}{\z@}%
                               {-3.5ex \@plus -1ex \@minus -.2ex}
                               {2.3ex \@plus.2ex}%
                               {\normalfont\large\bfseries}}
\renewcommand\subsection{\@startsection{subsection}{2}{\z@}%
                                 {-3.25ex\@plus -1ex \@minus -.2ex}%
                                 {1.5ex \@plus .2ex}%
                                 {\normalfont\bfseries}}
\makeatother

\numberwithin{equation}{section}

\def\Im{{\rm Im ~}}

\def\]{\right]}
\def\[{\left[}

\def\({\left (}
\def\){\right )}
\newcommand{\Tr}{{\rm Tr}}

\usepackage{wasysym }
\newcommand{\be}{\begin{equation}}
\newcommand{\ee}{\end{equation}}
\newcommand{\bea}{\begin{eqnarray}}
\newcommand{\eea}{\end{eqnarray}}
\newcommand{\ba}{\begin{eqnarray}}
\newcommand{\ea}{\end{eqnarray}}
\newcommand{\beal}{\begin{align}}
\newcommand{\eeal}{\end{align}}

\hypersetup{
    colorlinks,
    citecolor=blue,
    filecolor=black,
    linkcolor=red,
    urlcolor=black
}

\newcommand{\eg}{{\it e.g.,}\ }

\begin{document} 

\begin{titlepage}

\begin{center}

\phantom{ }
\vspace{3cm}

{\bf \huge{Long-distance $N$-partite\\  information for fermionic CFTs}}
\vskip 0.7cm
{\bf C\'esar A. Ag\'on\,${}^{\text{\neptune}}$, Pablo Bueno\,${}^{\text{\jupiter}}$ and Guido van der Velde\,${}^{\text{\jupiter}}$}
\vskip 0.75cm

\small{
\vskip -0.04in
${}^{\text{\neptune}}$\textit{Institute for Theoretical Physics, Utrecht University, 3584 CC Utrecht, The Netherlands}}
\vskip 0.3cm

\small{
\vskip -0.04in
${}^{\text{\jupiter}}$\textit{Departament de F{\'\i}sica Qu\`antica i Astrof\'{\i}sica,  Institut de Ci\`encies del Cosmos\\
 Universitat de
Barcelona,~Mart\'{\i} i Franqu\`es 1, E-08028 Barcelona, Spain}}

\vspace{0.4cm}
\textcolor{blue}{c.a.agonquintero@uu.nl,}
\textcolor{blue}{ pablobueno@ub.edu,}
\textcolor{blue}{ guidogvandervelde@gmail.com}

\begin{abstract}
The mutual information, $I_2$, of general spacetime regions is expected to capture the full data of any conformal field theory (CFT). For spherical regions, this data can be accessed from long-distance expansions of the mutual information of pairs of regions as well as of suitably chosen linear combinations of mutual informations involving more than two regions and their unions ---namely, the $N$-partite information, $I_N$.
In particular, the leading term in the $I_2$ long-distance expansion  
is fully determined by the spin and conformal dimension of the lowest-dimensional primary of the theory. When the operator is a scalar, an analogous formula for the tripartite information $I_3$ contains information about the OPE coefficient controlling the fusion of such operator into its conformal family. When it is a fermionic field, the coefficient of the leading term in $I_3$ vanishes instead. In this paper we present an explicit general formula for the long-distance four-partite information $I_4$ of general CFTs whose lowest-dimensional operator is a fermion $\psi$. The result involves a combination of four-point and two-point functions of $\psi$ and $\bar \psi$ evaluated at the locations of the regions. We perform explicit checks of the formula for a $(2+1)$-dimensional free fermion in the lattice finding perfect agreement. The generalization of our result to the $N$-partite information (for arbitrary $N$) is also discussed. Similarly to $I_3$, we argue that $I_5$ vanishes identically at leading order for general fermionic theories, while the $I_N$ with $N=7,9, \dots$ only vanish when the theory is free.  
\end{abstract}

\end{center}
\end{titlepage}
\setcounter{tocdepth}{2}
{\parskip = .2\baselineskip \tableofcontents}

\newpage
\section{Introduction}
The algebraic approach to Quantum Field Theory (QFT)  provides a universal description of high-energy physics in terms of assignations of operator algebras to spacetime regions \cite{Haag:1963dh}. Analogously to vacuum expectation values of quasi-local operators in the usual formulation \cite{Wightman:1956zz}, one can hope to describe a QFT in this context in terms of numbers obtained by acting on the algebras with the vacuum state. Such numbers would measure statistical properties of the vacuum restricted to the degrees of freedom attached to the regions. Notions borrowed from quantum information 
provide well suited candidates and, 
amongst those, the entanglement entropy (EE) stands out as the most natural choice. However, the EE of spacetime bipartitions is an ill-defined quantity in the continuum due to divergent correlations between fluctuations localized arbitrarily close to the entangling surface. One possibility is to introduce a UV regulator and extract the ``universal terms'' of the EE in a series expansion. Such terms turn out to contain a remarkable amount of information about the QFT. This includes renormalization group charges, trace-anomaly coefficients, stress-tensor correlators, thermal entropy coefficients, among others ---see \eg \cite{Calabrese:2004eu,Calabrese:2009qy,Casini:2011kv, Casini:2015woa,Holzhey_1994, Solodukhin:2008dh, Fursaev_2013,Safdi:2012sn,Miao:2015iba,Dowker:2010yj,Perlmutter:2013gua,Faulkner:2015csl,Bueno:2015rda,Bueno:2015qya,Swingle:2013hga,Bueno:2018xqc,Lee:2014xwa,Lewkowycz:2014jia,Anastasiou:2022pzm,Baiguera:2022sao,Bueno:2022jbl}. Alternatively, one can consider other quantities which are well-defined in the continuum. This naturally leads to the mutual information (MI) of pairs of regions $A,B$, which can be defined as
\begin{equation}
I_2(A,B)\equiv S(A)+S(B)-S(A B)\, ,
\end{equation}
where $S(A)$ is the EE of region $A$ with respect to its complement and $AB \equiv A \cup B$. Although the RHS must be computed in the presence of a UV regulator, the combination of EEs is such that divergences always cancel out and the MI remains finite and universal. 
 The MI satisfies several interesting properties \cite{Agon:2021zvp} such as being: positive semi-definite, $I_2(A,B) \geq  0$; symmetric in its arguments, $I_2(A,B)=I_2(B,A)$; and monotonous  under inclusions, $I_2(A,BC) \geq I_2(A,B)$.

The philosophy behind this entanglement formulation is that, in principle, knowledge of MI for all pairs of regions should be enough to uniquely reconstruct the underlying QFT model. In this regard, two regimes are specially useful to probe the theory. On the one hand, for concentric regions separated a short distance, the MI provides a geometric regulator for the EE which robustly captures the corresponding universal terms \cite{Casini:2007dk,Casini:2008wt,Casini:2009sr,Casini:2014yca,Bueno:2021fxb,Bueno:2023gey}.
Additionally, the short-distance regime also captures the phases of generalized symmetries \cite{Casini:2019kex, Casini:2020rgj}. 
On the other hand, when the regions are far apart and the theory is conformal, MI decays as inverse powers of the distance, the exponents being linear combinations of the conformal dimensions \cite{Cardy:2013nua}.  More specifically, when the regions are spherical, the long-distance expansion of the MI can be organized in terms of the conformal blocks associated to each primary operator \cite{Long:2016vkg, Chen:2016mya, Chen:2017hbk}. The leading term comes from the module with lightest weight $\Delta$, and reads
\begin{equation}\label{I2_intro}
I_2=	\#_{J}\,  c(\Delta) \frac{R_A^{2\Delta}R_B^{2\Delta}}{r^{4\Delta}}+
\dots \, ,\quad \text{where} \quad c(\Delta)\equiv \frac{\sqrt{\pi}}{4}\frac{\Gamma(2\Delta +1)}{\Gamma(2\Delta +\frac{3}{2})} \, . 
\end{equation} 
Here $R_{A,B}$ are the radii of the spheres and $r$ is the separation between their centers. The dots represent subleading terms in the $R_A R_B/r$ expansion. On the other hand, the coefficient $\#_J$ depends on the spin $J$ of the lowest lying primary $\mathcal{O}$, and involves a contraction between the timelike normal vector to the spheres $n_{A,B}$ and the unit vector $\hat{r}$. For instance, for spin-$0$ and spin-$1/2$ fields, one finds, respectively  \cite{Calabrese:2010he,Agon:2015ftl,Casini:2021raa},
\begin{eqnarray}
\#_{J=0} & = &  1 \, ,\\\label{c_fermion_intro}
\#_{J=\frac{1}{2}}& = & 2^{\left[ \frac{d}{2}\right]+1}\left[ 2(n_A \cdot \hat{r})(n_B \cdot \hat{r})-(n_A\cdot n_B)\right]\, .
\end{eqnarray}

Information about the complete spectrum as well as the OPE coefficients, which is all we need to reconstruct the CFT, should be accessible from the subleading terms in the MI long-distance expansion. However, identifying these terms in practice is in general very challenging ---see \cite{Agon:2021zvp} for an example in the case of a $d$-dimensional free fermion. Alternatively, we can perform a similar long-distance analysis for entanglement measures constructed from linear combinations of mutual informations but which involve a greater number of entangling regions. The simplest case corresponds to the tripartite information.
This is defined for three regions $A$, $B$, $C$, as
\begin{eqnarray}
I_3(A,B,C)&\equiv &I_2(A,B)+I_2(A,C)-I(A,BC)\\
&=& S(A)+S(B)+S(C)-S(AB)-S(AC)-S(BC)+S(ABC)\, .
\end{eqnarray}
Tripartite information measures the extensivity of mutual information, and unlike the latter, has no definite sign. While holographic theories are said to be ``monogamous'', with $I_3 \leq 0$ for arbitrary regions \cite{Hayden:2011ag,  Cui:2018dyq}, free models like the scalar and the Dirac field in dimension grater than two feature $I_3>0$ \cite{Casini:2008wt}.   The case $I_3=0$ for general regions is only possible for a two-dimensional free fermion \cite{Agon:2021zvp}, and has motivated the so-called Extensive Mutual Information (EMI) model \cite{Casini:2008wt}. 
When the primary with lowest dimension is a scalar field and the regions are spherical, it was shown in \cite{Agon:2021lus} that the leading long-distance term of the tripartite information is given by
\begin{equation}
I_3=\left[ \frac{2^{6\Delta}\Gamma(\Delta +\frac{1}{2})^3}{2\pi \Gamma(3\Delta+\frac{3}{2})}-c\left(\frac{3}{2}\Delta\right)(C_\mathcal{OOO})^2\right]\frac{R_A^{2\Delta}R_B^{2\Delta}R_C^{2\Delta}}{r_{AB}^{2\Delta}r_{AC}^{2\Delta}r_{BC}^{2\Delta}}+
\dots \, , \quad (J=0)  
\end{equation}
which holds for $R_{A,B,C}\ll r_{AB},r_{AC},r_{BC}$. Here,
 $C_\mathcal{OOO}$ is the OPE coefficient giving the fusion of the lowest-dimensional primary $\mathcal{O}$ into its conformal family. As explained in \cite{Agon:2021lus}, only for large values of  $C_\mathcal{OOO}$ the MI is monogamous at long distances. If the lowest-dimensional primary is a fermion, on the other hand, the analogous leading term for the tripartite information identically vanishes,
 \begin{equation}\label{tripav}
I_3=0+\dots \, , \quad (J=1/2)
\end{equation}
implying a scaling with the inverse distance to a power grater than $6\Delta$. In that case, accessing OPE coefficients or more refined CFT data requires moving to subleading orders or considering a generalized measure involving even more regions.

The obvious candidate is the $N$-partite information which, given $N$ disjoint regions $A_1, \ldots , A_N \equiv \{A_i\}$, is defined as
\begin{equation}\label{IN_definition}
I_N(\{A_i\})=\sum_i S(A_i)-\sum_{i<j}S(A_i A_j)+\sum_{i<j<k}S(A_iA_jA_k)-\cdots + (-1)^{N+1}S(A_1\cdots A_N)\, , 
\end{equation}
or, equivalently, as
\begin{equation}
I_N(\cdot,A_{N-1},A_N)=I_{N-1}(\cdot,A_{N-1})+I_{N-1}(\cdot,A_{N-1})-I_{N-1}(\cdot, A_{N-1}A_N)\, .
\end{equation}
where $\cdot \equiv A_1,   \ldots, A_{N-2}$, which manifestly shows that $I_N$ is a measure of the extensivity of $I_{N-1}$. Naturally, $I_N$ can also be written as a linear combination of mutual informations involving various unions of regions. The literature involving studies of the $N$-partite information for $N\geq 4$ in the context of quantum field theory is rather limited and has been mostly confined to holographic theories \cite{Hayden:2011ag,Alishahiha:2014jxa,Mirabi:2016elb,Bao:2015bfa,Hubeny:2018trv,Hubeny:2018ijt,He:2019ttu,HernandezCuenca:2019wgh,Fadel:2021urx} and two-dimensional CFTs \cite{Coser:2013qda,DeNobili:2015dla}. An exception is \cite{Agon:2022efa}, where it was argued that the $N$-partite information of spacetime regions in a general $d$-dimensional CFTs can be expressed as the expectation value of $N$ twist operators implementing the identification of the replica sheets along the entangling regions. Using this, it was shown that in the long-distance limit it behaves as
\begin{equation}
    I_N(\{A_i\}) \propto \left[\frac{R}{r} \right]^{2N\Delta}+\dots \, \quad (J=0)\, ,
\end{equation}
where $\Delta$ is the lowest primary dimension and for simplicity it was assumed that all spheres have equal radius $R$ and are separated a distance of the same order $r$. Furthermore, it was argued that this leading term in $I_N$ includes $N$-,$(N-1)$-,$\dots$ and 2-point correlators of the smallest-dimension primary operator. Hence, the $I_N$ long-distance expansion provides an alternative route for extracting the CFT data.

In this paper, we generalize the results of \cite{Agon:2022efa} to the case in which the lowest-dimensional operator is a spin-$1/2$ field. Considering the long-distance regime, we write the twists as the OPE of spinors supported on different sheets. We are left with a linear combination of products of as many correlation functions as replica indices with non-trivial support. We follow the diagrammatic representation of \cite{Agon:2022efa}, adapted to the fermionic case, to account for the different contributions to multipartite information.  Armed with this graph technology, we compute the general formula for the leading term in the long-distance expansion of the four-partite information, valid for CFTs with a spin-$1/2$ field as their lowest-dimensional primary. The result reads
\begin{equation}
I_4=\left(R_A R_B R_C R_D\right)^{2\Delta} n_A^{\mu} n_B^{\nu}n_C^{\lambda}n_D^{\eta} \left(\gamma_{\mu}\right)_{\alpha\beta}\left(\gamma_{\nu}\right)_{\rho\sigma}\left(\gamma_{\lambda}\right)_{\pi\zeta}\left(\gamma_{\eta}\right)_{\theta\tau}\mathcal{T}+\dots\, ,
\end{equation}
where we label the regions by $\{A,B,C,D\}$ and, schematically
\begin{align}
\mathcal{T}=&+c_1 \left[\left\langle \bar{\psi}_A\bar{\psi}_B \psi_C\psi_D\right\rangle_{\rm conn} \left\langle \psi_A\psi_B\bar{\psi}_C\bar{\psi}_D\right\rangle_{\rm conn} +(B\leftrightarrow \{C, D\})\right]
\\ \notag &+(c_3-c_1) \left[\left\langle \bar{\psi}_A\psi_D\right\rangle \left\langle \psi_A \bar{\psi}_B\right\rangle \left\langle \psi_B\bar{\psi}_C \right\rangle \left\langle \psi_C \bar{\psi}_D\right\rangle +\text{permutations of}~\{ B,C,D\}\right]\, .
\end{align}
Here $\psi_A$ means that the spinor is evaluated at $r_A$ and a spinor index is also implicit. Meanwhile, $c_1$ and $c_3$ are numerical coefficients defined in (\ref{c1def}) and (\ref{c3def}) below. What we would like to stress now is that this formula depends not only on the spinor two-point function, but also on its (connected) four-point function. Hence, as opposed to the $I_2$ and $I_3$, from the $I_4$ long-distance leading term we can access the structure constants appearing in the conformal block decomposition.


The four-partite information formula simplifies significantly when the theory is free and the spheres are situated at a fixed time slice $t=0$. It reads
\begin{equation}\label{I4_free0}
\begin{split}
 I_4 =\, \,& c_{\text{free}}\, 2^{[(d+2)/2]} \, (R_A R_B R_C R_D)^{d-1}\times  \\
&\Bigg\{ \frac{\left[ \left(\hat{r}_{AB}\cdot \hat{r}_{AD}\right)\left(\hat{r}_{BC}\cdot \hat{r}_{CD}\right)+\left(\hat{r}_{AB}\cdot \hat{r}_{BC}\right)\left(\hat{r}_{AD}\cdot \hat{r}_{CD}\right)-\left(\hat{r}_{AB}\cdot \hat{r}_{CD}\right)\left(\hat{r}_{AD}\cdot \hat{r}_{BC}\right)\right]}{ (\vert r_{AB}\vert\vert r_{AD}\vert\vert r_{BC}\vert\vert r_{CD}\vert)^{d-1} }\\
&\, \, -  \frac{\left[ \left(\hat{r}_{AB}\cdot \hat{r}_{AC}\right)\left(\hat{r}_{BD}\cdot \hat{r}_{CD}\right)+\left(\hat{r}_{AB}\cdot \hat{r}_{BD}\right)\left(\hat{r}_{AC}\cdot \hat{r}_{CD}\right)-\left(\hat{r}_{AB}\cdot \hat{r}_{CD}\right)\left(\hat{r}_{AC}\cdot \hat{r}_{BD}\right)\right]}{(\vert r_{AB}\vert\vert r_{AC}\vert\vert r_{BD}\vert\vert r_{CD}\vert)^{d-1}}\\
&\, \, - \frac{\left[ \left(\hat{r}_{AC}\cdot \hat{r}_{AD}\right)\left(\hat{r}_{BC}\cdot \hat{r}_{BD}\right)+\left(\hat{r}_{AC}\cdot \hat{r}_{BC}\right)\left(\hat{r}_{AD}\cdot \hat{r}_{BD}\right)-\left(\hat{r}_{AC}\cdot \hat{r}_{BD}\right)\left(\hat{r}_{AD}\cdot \hat{r}_{BC}\right)\right]}{(\vert r_{AC}\vert\vert r_{AD}\vert\vert r_{BC}\vert\vert r_{BD}\vert)^{d-1}} \Bigg\}\, .
\end{split}
\end{equation}
Here, $d$ is the spacetime dimension, $c_{\text{free}}$ is another dimension-dependent constant defined in (\ref{c_free_d}), and $\hat{r}_{AB}\equiv (r_A-r_B)/\vert r_A-r_B\vert$. We compare this analytic expression for the $(2+1)$-dimensional free fermion with numerical results obtained in the lattice. We consider 
two different geometrical arrangements, and find a perfect agreement in both cases. Additionally, from our general formula (\ref{I4_free0}) we are able to show that the $(2+1)$-dimensional free-fermion $I_4$ does not have a definite sign, namely, that there exist geometric configurations for which $I_4$ is positive, negative and zero.

Regarding multi-partite information for a larger number of regions, we are able to prove that $I_5=0$ at leading order for general CFTs whose lightest primary is a Dirac spinor. In turn, we show that $I_N$ with odd $N\geq 7$ also vanishes at leading order, but only as long as the theory is free. 

The plan of the remainder of the article is as follows. In section \ref{sec:general_structure} we go over the most important steps in the derivation of the formula for the $N$-partite information as an expectation value of a product of $N$ twist operators. The fact that the regions are far apart allows us to approximate the twists as a bilinear in the lightest primary spinor, being antisymmetric in the replica indices and evaluated at a conveniently chosen location inside the corresponding region. Moreover, we derive the expression for the coefficients of this expansion, which we call $b_{ij}$. 
The derivation follows \cite{Agon:2022efa} and relies heavily on the spherical symmetry of the entangling surfaces, since in that case the modular evolution is a well-known conformal transformation. In sections \ref{sec:mutual} and \ref{sec:tripartite} we review the computations of the corresponding leading terms for the mutual and tripartite informations, respectively. Applying the graph representation to organize the different contributions, we recover (\ref{I2_intro}), (\ref{c_fermion_intro}) and (\ref{tripav}), as expected. The four-partite information is dealt with in section \ref{sec:fourpartite}. There, we manage to express the $I_4$ as a linear combination of products of fermionic two-point functions and connected four-point functions. The coefficients are equal to the contraction of four $b_{ij}$, and we are able to calculate them by adapting the bosonic formulae in \cite{Agon:2022efa} rather straightforwardly. Later, we focus on the theory of a free fermion, where the expression simplifies because the connected four-point function vanishes. We compare the analytic prediction with numerical results obtained in the lattice in the $d=3$ case, corresponding to configurations where the spheres are placed either in the vertices of a square or within a line, finding excellent agreement in both cases. Also, in order to study the possible signs of $I_4$, we rewrite it in terms of a few independent geometrical parameters, finding configurations where it takes either negative, zero, or positive values. At the end of section \ref{sec:fourpartite} we comment on the general structure of the $I_4$ for interacting fermionic CFTs. 
In section \ref{sec:N_parite}, we analyze multipartite information for $N>4$. We show that, analogously to the $I_3$, the leading term in the $I_5$ identically vanishes for theories with a fermionic lowest-dimensional operator. However, we argue that this trend continues for $N=7,9,\dots$ only when the theory is free.  We end this article in section \ref{sec:final_comments} with some final remarks. We leave for appendices \ref{charge_conj} and \ref{fmc} some digression on the charge conjugation invariance of the fermionic twist operator and the fermionic modular correlator, respectively.

\section{$N$-partite information: general structure}\label{sec:general_structure}
In this section we review the derivation of the formula for the long-distance leading contribution to the $N$-partite information in terms of an expectation value of $N$ twist operators with support on the corresponding entangling regions \cite{Agon:2022efa}. We apply the formula to the case in which the lowest-scaling dimension of the CFT is a fermionic field.

The first step is considering the R\'enyi $N$-partite information, which replaces every entanglement entropy in (\ref{IN_definition}) by a R\'enyi entropy of index $n$. We shall take the $n\rightarrow 1$ limit to recover the $N$-partite information at the end. 
This procedure turns out to be useful since it allows us to express $I_N$ as an expectation value of a product of twist operators implementing the conical singularity characterizing the replica manifold. When the regions are very far from each other, the sewing of the different copies of the field taking place along the entangling surfaces can be effectively approximated by the fusion of local operators at some conveniently chosen points inside the regions \cite{Cardy:2013nua}, leading to ordinary correlation functions. 

Let us be more explicit. For a given region $A$, we use the replica trick to express its vacuum R\'enyi entropy $S^{(n)}(A)$ in terms of the partition function on the replica manifold $Z(\mathcal{C}^{(n)}_A)$,
\begin{equation}\label{renyi1}
S^{(n)}(A)=\frac{1}{1-n}\log{\left[\frac{Z(\mathcal{C}^{(n)}_A)}{Z^n}\right]}\, .
\end{equation}
In turn, the above partition function can be alternatively written as the expectation value of a twist operator $\Sigma^{(n)}_A$ on $n$ copies of the original manifold $\mathcal{M}^{(n)}$ \cite{Calabrese:2004eu,Calabrese:2005zw,Hung:2014npa}
\begin{equation}\label{twist_Z}
\left\langle\Sigma^{(n)}_A\right\rangle=\frac{Z(\mathcal{C}^{(n)}_A)}{Z^n}\, .
\end{equation}
The twist operator is a non-local operator with support on the entangling region $A$ that implements the identification of the $i$-th copy with the $i+1$-th one. Anticipating the $n\rightarrow 1$ limit, we normalize the twist operator as follows,
\begin{equation} \label{twist1}
\Sigma^{(n)}_A=\left\langle\Sigma^{(n)}_A\right\rangle \left( 1+ \tilde{\Sigma}^{(n)}_A\right)\, ,
\end{equation}
such that $\left\langle \tilde{\Sigma}^{(n)}_A\right\rangle=0$. Substituting (\ref{renyi1}), (\ref{twist_Z}) and (\ref{twist1}) into (\ref{IN_definition}), it turns out that the only contribution to the $N$-partite information is \cite{Agon:2022efa}
\begin{equation}\label{I_N}
I_N(\{A_i\})=\lim_{n\rightarrow 1}\frac{(-1)^{N+1}}{1-n}\left\langle \tilde{\Sigma}_{A_1}^{(n)}\tilde{\Sigma}_{A_2}^{(n)}\cdots\tilde{\Sigma}_{A_N}^{(n)} \right\rangle\, .
\end{equation}
At long distances the twist operator $\tilde{\Sigma}_{A}^{(n)}$ can be approximated by the operator product expansion of local operators in each sheet. The leading contribution stems from the product of a pair of operators with the lowest dimensions of the theory.  For a theory with a fermionic field $\psi$ with scaling dimension $\Delta$ as the lowest-lying primary, its explicit form reads
\cite{Chen:2017hbk,Casini:2021raa,Agon:2021zvp}
\begin{equation}\label{twist_OPE}
\tilde{\Sigma}_{A}^{(n)}\approx \sum_{i\neq j} \frac{1}{2}b_{ij}^A n_{A,\mu}\left(\bar{\psi}^i (r_A)\gamma^{\mu}\psi^j(r_A)-\bar{\psi}^j (r_A)\gamma^{\mu}\psi^i(r_A)\right) \, ,
\end{equation}
where $i,j$ are replica indices, $n_A$ is the vector normal to the region and more will be said about the coefficients $b_{ij}^A$ below. Terms of the form $\bar{\psi}\psi$ do not contribute for a massless field due to chiral symmetry in even dimensions or parity in odd dimensions \cite{Casini:2021raa}. Moreover, antisymmetric tensors $\bar{\psi}\left[\gamma^\mu\gamma^\nu \right]\psi$ vanish as well because there is only one available vector $n_A$ to contract with. Note also that the bilinear is antisymmetric in the replica indices. This is required by charge conjugation invariance, as we show in appendix \ref{charge_conj}.

Plugging the OPE expansion (\ref{twist_OPE}) back in (\ref{I_N}), we get
\begin{equation}\label{IN_final}
\begin{split}
I_N(\{A_i\})=&(-1)^{N+1} (n_{A_1})_{\mu_1}\cdots (n_{A_N})_{\mu_N}\gamma^{\mu_1}_{a_1b_1}\cdots \gamma^{\mu_N}_{a_Nb_N}\\
& \lim_{n\rightarrow 1} \frac{1}{1-n} \sum_{i_1\neq j_1}\cdots \sum_{i_N \neq j_N}b_{[i_1j_1]}^{A_1}\cdots b_{[i_Nj_N]}^{A_N}\left\langle \bar{\psi}^{i_1}_{a_1} (r_1)\psi^{j_1}_{b_1}(r_1)\cdots \bar{\psi}^{i_N}_{a_N} (r_N)\psi^{j_N}_{b_N}(r_N)\right\rangle
\end{split}
\end{equation}
with latin indices $\{a_i,b_i\}$ labeling spinor components. This expectation value factorizes into the product of as many expectation values as the number of sheets $\mathcal{N}$ with non trivial operator insertions, because spinors at different sheets are uncorrelated. Therefore, the fact that the leading long-distance behavior of the $N$-partite information depends on the number of distinct sheets suggests a graph representation \cite{Agon:2022efa}, where vertices represent the different copies and the arrows connecting them stand for the region on which the bilinear with such replica indices is located. Note that each expectation value should have an equal number of spinors and conjugated spinors, so if we conventionally represent an arrow flowing towards a vertex as a $\psi$ and an arrow leaving it as $\bar{\psi}$, then the condition that these must be balanced within a correlation function would mean that the total flux to each vertex should be zero. In the following sections we will show how this diagrammatic representation helps identifying and organizing the different contributions in a simple way in the particular cases of $N=2,3, 4$.

\subsection{The $b_{ij}$ coefficients}
Having expressed the $N$-partite information in terms of the lightest primary correlation functions, what we need to compute next are the $b_{ij}$ coefficients giving the long distance expansion of the twist operator (\ref{twist_OPE}). For reasons that will become apparent soon, we focus on the case of spherical regions. In that situation, we can factor out the radius $R_A$, which is the only characteristic scale, and define
\begin{equation}\label{b}
b_{i j}^A=b_{i j} R_A^{2\Delta}\, .
\end{equation}
so that the new coefficients $b_{i j}$ are scale invariant. In turn, these can be read off from the two point function 
\begin{equation}\label{corr_with_twist}
\bar{n}^\mu\left\langle\tilde{\Sigma}_{A}^{(n)}\bar{\psi}_{\lambda}^i(x)\left(\gamma_{\mu}\right)_{\lambda\rho}\psi_{\rho}^j(x)\right\rangle \, ,   
\end{equation}
where $\bar{n}^\mu$ is an arbitrary future directed normal time-like vector and $\vert x-x_A\vert^2\rightarrow \infty$. In fact, using (\ref{twist_OPE}) and the primary spinor correlator
\begin{equation}\label{2pointfunction}
\left\langle \psi_\alpha(r_A)\bar{\psi}_\beta(r_B)\right\rangle=i\left(\gamma^{\mu}\right)_{\alpha\beta}\frac{(r_B-r_A)_\mu}{\vert r_B-r_A\vert^{2\Delta+1}}\, ,    
\end{equation}
we arrive at \cite{Casini:2021raa,Agon:2021zvp}
\begin{equation}\label{coeff}
b_{[i j]}=\lim_{x\rightarrow \infty}2^{-\left[\frac{d}{2}\right]}\frac{x^{4\Delta}}{R_A^{2\Delta}}\frac{\bar{n}^{\mu}\left\langle \tilde{\Sigma}_{A}^{(n)}\bar{\psi}_{\lambda}^i(x)\left(\gamma_{\mu}\right)_{\lambda\rho}\psi_{\rho}^j(x)\right\rangle}{\left[2\left(n_A\cdot \hat{x}\right)\left(\bar{n}\cdot \hat{x}\right)-\left(n_A \cdot \bar{n}\right)\right]}\, ,
\end{equation}
where $b_{[i,j]}=(b_{ij}-b_{ji})/2$. 
Furthermore, (\ref{corr_with_twist}) is related to the correlation between the spinor and its modular transformed version \cite{Casini:2021raa}. 
\begin{equation}\label{numerator}
\bar{n}^\mu\left\langle \tilde{\Sigma}_{A}^{(n)}\bar{\psi}_{\lambda}^i(x)\left(\gamma_{\mu}\right)_{\lambda\rho}\psi_{\rho}^j(x)\right\rangle \underset{n\to 1}{=} \left\{\begin{array}{ll}
\!\!\!
- \slashed{\bar{n}}_{\lambda\rho} \left\langle \bar{\psi}_\lambda (x) \psi_\rho(x,i\tau_{i j}+i)\right\rangle \,, &{\rm for}\quad  \!\!i<j\\
\,\,\, \slashed{\bar{n}}_{\lambda\rho} \left\langle \bar{\psi}_\lambda (x) \psi_\rho(x,i \tau_{i j})\right\rangle \,, &{\rm for}\quad  \!\!i>j
\end{array} \right. 
\end{equation}
where $\tau_{ij}= (i-j)/n$ and 
\begin{equation}
\psi(x,s)\equiv \rho_A^{-i s} \psi(x) \rho_A^{i s}\, .
\end{equation}
Terms of order $\mathcal{O}(n-1)^2$ are dismissed since these will be subleading in the $n\rightarrow 1$ limit defining the $N$-partite information. Equation (\ref{numerator}) is a bit subtle in the fermionic case due to the antisymmetric nature of the spinors and the KMS condition. We present a derivation in Appendix \ref{fmc}. Crucially, when region $A$ is a ball and $0<\Im(s)<1$, we have an analytic expression for the modular flow, which is the conformal map\footnote{We assume that the spherical region is at $x^0=0$, or equivalently, that $n=\hat{t}$.} \cite{Casini:2021raa}
\begin{equation}
\begin{split}
x^0(s)&=N(s)^{-1}R\left(x^0 R\cosh(2\pi s)+\frac{1}{2}\left(R^2-x^2\right)\sinh(2\pi s)\right)\, ,\\
x^i(s)&=N(s)^{-1}R^2 x^i\, , \\
N(s)&= x^0 R \sinh(2\pi s)+\frac{1}{2}\cosh(2\pi s) \left(R^2-x^2\right)+\frac{1}{2}\left(R^2+x^2\right)\, .
\end{split}
\end{equation}
with $x^2=-(x^0)^2+x^i x^i$. In the limit of large distance $\vert x \vert\rightarrow \infty$ the associated coordinate transformation matrix is
\begin{equation}
\frac{d x^\mu}{d x^\nu}=\Omega(x,s)\Lambda^\mu_\nu (x,s)\, \quad{\rm where }\quad \label{Omega}
\Omega(x,s)\approx \frac{-R^2}{x^2 \sinh^2{(\pi s)}}
\end{equation}
holds in that limit. The Lorentz transformation is the boost with parameter  
\begin{equation}
\cosh{\beta}=-\frac{{x^0}^2+{x^i}^2}{x^2}\, \quad \quad
\sinh{\beta}=-\frac{2 x^0 \vert x^i \vert}{x^2}\,,
\end{equation}
in the direction $\hat{x}^i$. 
Hence, the fermion transforms as
\begin{equation}\label{conformalSpinor}
\psi_\rho(x,s)=\Omega(x,s)^{\Delta} S_\rho^{\sigma}\left(\Lambda^{-1}(x,s)\right)\psi_\sigma (x(s))\, ,
\end{equation}
where
\begin{eqnarray}
S\left(\Lambda^{-1}(x,s)\right)= \cosh{\frac{\beta}{2}}-\gamma^0\gamma^i \hat{x}_i \sinh{\frac{\beta}{2}}=-i\slashed{\hat{x}}\slashed{n}\,. \label{lorentzSpinor}
\end{eqnarray}
Moreover, in the $\vert x \vert \rightarrow \infty$ limit
\begin{eqnarray}\label{x_s}
x^0(s)&\approx & R \coth{\pi s} - x^0 \Omega(s)\,,\nonumber \\
x^i(s)&\approx &  x^i \Omega(s).
\end{eqnarray}
Substituting (\ref{conformalSpinor}), (\ref{Omega}), (\ref{lorentzSpinor}) and (\ref{x_s}) into (\ref{numerator}), we get
\begin{equation}
\bar{n}^\mu\left\langle \tilde{\Sigma}_{A}^{(n)}\bar{\psi}_{\lambda}^i(x)\left(\gamma_{\mu}\right)_{\lambda\rho}\psi_{\rho}^j(x)\right\rangle \approx 2^{\left[\frac{d}{2}\right]}\frac{R^{2\Delta} {\rm sgn}(i-j)}{x^{4\Delta}\sin^{2\Delta}{\left[\pi |\tau_{ij}|\right]}}\left[2\left(n\cdot\hat{x}\right)\left(\bar{n}\cdot\hat{x}\right)-\left(\bar{n}\cdot n\right)\right]\, .
\end{equation}
where the sign function above allows to express equation (\ref{numerator}) which is defined by parts with a single function for all $i,j$. 
Finally, substitution into (\ref{coeff}) leads to 
\begin{equation}\label{b_final}
b_{[ij]}=\frac{\text{sgn}(i-j)}{\sin^{2\Delta}{\left[\pi |\tau_{ij}|\right]}}\,.
\end{equation}
Note that the absolute value of (\ref{b_final}) coincides with the coefficients found when the primary with the lowest scaling dimension is a scalar\cite{Agon:2022efa}. As we show in the next sections, this will allow us to relate straightforwardly the sums involved in (\ref{I_N}) to the ones computed in \cite{Agon:2022efa}.

\section{Mutual information}\label{sec:mutual}
As a warm up, in this section and the following we rederive the previously known results of the mutual and tripartite informations of two and three spheres, respectively, in the long-distance regime for CFTs whose lowest-scaling operator is a spin-$1/2$ field. 

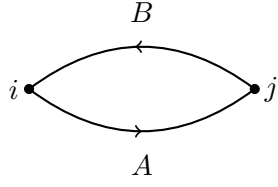
\begin{figure}[t!]
\centering
\begin{tikzpicture}
\begin{scope}[thick,decoration={
    markings,
    mark=at position 0.5 with {\arrow{>}}}
    ] 
\draw [postaction={decorate}](-1.5,0) .. controls (-0.5,-0.75) and (0.5,-0.75) .. (1.5,0);
\end{scope}
\begin{scope}[thick,decoration={
    markings,
    mark=at position 0.5 with {\arrow{<}}}
    ] 
\draw [postaction={decorate}](-1.5,0) .. controls (-0.5,0.75) and (0.5,0.75).. (1.5,0);
  \end{scope}
\draw [fill] (-1.5,0) circle[radius=0.06]; 
\draw [fill] (1.5,0) circle[radius=0.06];

\node [right] at (1.5,0) {$j$};
\node [left] at (-1.5,0) {$i$};
\node [below] at (0,-0.75) {$A$};
\node [above] at (0,0.75) {$B$};
\end{tikzpicture}
\captionsetup{width=0.9\textwidth}
\caption{Diagram representing the only configuration that contributes to $I_2$, where two replica sheets are different.}
\label{I2diagram}  
\end{figure}

We start here with the $I_2$, that is, we
evaluate (\ref{IN_final}) in the particular case of $N=2$. If we label the entangling regions as $A$ and $B$, then\footnote{From now on $b_{ij}$ must be understood as its antisymmetric part $b_{[i j]}$.}
\begin{equation}\label{I2}
I_2= \left(R_A R_B\right)^{2\Delta} n_A^{\mu} n_B^{\nu} \left(\gamma_{\mu}\right)_{\alpha\beta}\left(\gamma_{\nu}\right)_{\rho\sigma}\left[ \lim_{n\rightarrow 1}\frac{1}{n-1}\sum_{i\neq j}\sum_{k\neq \ell}b_{ij}b_{k\ell}\left\langle \bar{\psi}_{\alpha}^i (r_A)\psi_{\beta}^j(r_A)\bar{\psi}_{\rho}^k (r_B)\psi_{\sigma}^\ell(r_B)\right\rangle \right]+\dots\, .
\end{equation}
There is a single contribution to the expression in square brackets, coming from the configuration with two different replica sheets. This is then represented by a graph with two vertices, linked by a couple of arrows flowing in opposite directions, as shown in Fig.\,\ref{I2diagram}.
More concretely, we have
\begin{eqnarray}
\left[~\right]&=&\lim_{n\rightarrow 1}\frac{1}{n-1}\sum_{i\neq j}b_{ij} b_{ji} \left\langle \bar{\psi}_{\alpha}^i (r_A)\psi_{\sigma}^i(r_B)\right\rangle\langle\psi_{\beta}^j (r_A)\bar{\psi}_{\rho}^j(r_B)\rangle\\
&=& \left(\lim_{n\rightarrow 1}\frac{1}{n-1}\sum_{i\neq j}b_{ij} b_{ji}\right)\left( -i (\gamma^{\theta})_{\sigma\alpha}\frac{(r_A-r_B)_\theta}{\vert r_A-r_B\vert^{2\Delta+1}}\right)\left(i (\gamma^{\eta})_{\beta\rho}\frac{(r_B-r_A)_\eta}{\vert r_B-r_A\vert^{2\Delta+1}}\right)\\
&=& \left(\lim_{n\rightarrow 1}\frac{1}{n-1}\sum_{i\neq j}b_{ij}b_{ji}\right) (-1) (\gamma^{\theta})_{\sigma\alpha}(\gamma^{\eta})_{\beta\rho}\frac{(\hat{r}_{AB})_\theta}{\vert r_{AB}\vert^{2\Delta}}\frac{(\hat{r}_{AB})_\eta}{\vert r_{AB}\vert^{2\Delta}}\, .
\end{eqnarray}
Following the conventions of \cite{Agon:2022efa}, we define
\begin{equation}
\left(\lim_{n\rightarrow 1}\frac{1}{n-1}\sum_{i\neq j}b_{ij} b_{j i}\right)\equiv - 2 c_{2:2}^{(2)}=-\frac{\sqrt{\pi}}{2}\frac{\Gamma(2\Delta+1)}{\Gamma(2\Delta+3/2)}\, .
\end{equation}
Substituting back in (\ref{I2}), and using $\hat{r}\equiv \hat{r}_{AB}$ for short,
\begin{eqnarray}
I_2&=& 2 c_{2:2}^{(2)} \frac{(R_A R_B)^{2\Delta}}{r^{4\Delta}}(n_A)_{\mu}(n_B)_{\nu}\hat{r}_\theta\hat{r}_\eta \text{Tr}\left[\gamma^{\mu}\gamma^{\eta}\gamma^{\nu}\gamma^{\theta}\right]\\
&=& 2 c_{2:2}^{(2)} \frac{(R_A R_B)^{2\Delta}}{r^{4\Delta}} 2^{[d/2]}\left[2(n_A\cdot\hat{r})(n_B\cdot\hat{r})-(n_A\cdot n_B)\right]\, ,
\end{eqnarray}
which leads to the final result
\begin{equation}
I_2=2^{[d/2]+1}\frac{\sqrt{\pi}}{4}\frac{\Gamma(2\Delta+1)}{\Gamma(2\Delta+3/2)}\frac{(R_A R_B)^{2\Delta}}{r^{4\Delta}}\left[2(n_A\cdot\hat{r})(n_B\cdot\hat{r})-(n_A\cdot n_B)\right]+\dots
\end{equation}
In this last expression we have introduced dots to denote subleading corrections in the size/distance ratio.
The formula matches the result of \cite{Chen:2017hbk, Casini:2021raa}, as expected. Notice that the effect of the leading-primary spin-$1/2$ appears through the non-trivial tensorial structure involving the normal vectors to the sphere planes and the vector which connects their centers. Analogous formulas for leading primaries transforming on more general representations of the Lorentz group can be found in \cite{Casini:2021raa}.

\section{Tripartite information}\label{sec:tripartite}
Consider now an additional ball, which we label with letter $C$. The tripartite information for the three balls separated by distances much greater than their sizes reads
\begin{eqnarray}\nonumber
I_3&=& \left(R_A R_B R_C\right)^{2\Delta}(n_{A})_\mu (n_{B})_\nu (n_{C})_\sigma \gamma^\mu_{a b}\gamma^\nu_{c d}\gamma^\sigma_{e f}\\\label{I3}
&~& \left[\lim_{n\rightarrow 1}\frac{1}{1-n}\sum_{i\neq j\neq k}b_{i j}b_{j k}b_{k i}\left\langle \bar{\psi}^i_a (r_A)\psi^j_b(r_A)\bar{\psi}^k_c (r_B) \psi^\ell_d(r_B)\bar{\psi}^r_e (r_C)\psi^s_f(r_C)\right\rangle\right]+\dots 
\end{eqnarray}
Current conservation implies that only graphs with $\mathcal{N}=3$ vertices contribute. However, the second line of (\ref{I3}) involves two different terms, corresponding to the two possible ways of labeling the free sides of the triangle in Fig.\,\ref{N3diagram_I3},
\begin{figure}[t!]
\centering
\begin{tikzpicture}
\begin{scope}[ thick,decoration={
    markings,
    mark=at position 0.5 with {\arrow{>}}}
    ] 
\draw [postaction={decorate}] (0,-2) --(1,0);
 \draw [postaction={decorate}] (1,0)-- (-1,0); 
 \draw [postaction={decorate}] (-1,0) --(0,-2);
  \end{scope}
\draw [fill] (0,-2) circle[radius=0.06]; 
\draw [fill] (1,0) circle[radius=0.06];
\draw [fill] (-1,0) circle[radius=0.06]; 
\node [below] at (0,-2) {$k$};
\node [right] at (1,0) {$j$};
\node [left] at (-1,0.1) {$i$};
\node [above] at (0,0) {$A$};
\end{tikzpicture}
\captionsetup{width=0.9\textwidth}
\caption{Diagram representing the two terms contributing to $I_3$, corresponding to the permutation of regions $B$ and $C$ associated to the free sides of the triangle.}
\label{N3diagram_I3} 
\end{figure}
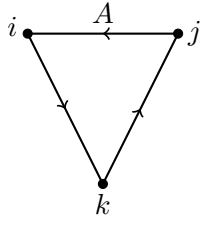 
\begin{equation}\label{I3_intermediate_step}
\begin{split}
\left[~\right]=&\left(\lim_{n\rightarrow 1}\frac{1}{1-n}\sum_{i\neq j\neq k}b_{i j}b_{j k}b_{k i}\right)\\
&\left[\left\langle \bar{\psi}^i_a(r_A) \psi^i_f(r_C)\right\rangle  \left\langle \psi^j_b(r_A) \bar{\psi}^j_c(r_B)\right\rangle \left\langle \psi^k_d(r_B) \bar{\psi}^k_e(r_C)\right\rangle + \left(B\leftrightarrow C\right) \right]\, ,
\end{split}
\end{equation}
where $\left(B\leftrightarrow C\right)$ means that we exchange the pair $\bar{\psi}_{B}\leftrightarrow \bar{\psi}_{C}$, $\psi_{B}\leftrightarrow \psi_{C}$, with their respective spinor indices. Note that this commutation keeps the overall sign unaltered because the bilinear $\bar{\psi}\psi$ is bosonic. Using (\ref{2pointfunction}) and substituting in (\ref{I3}) we get
\begin{equation}
\begin{split}
I_3=&i \left( R_A R_B R_C\right)^{2\Delta}\left(\lim_{n\rightarrow 1}\frac{1}{1-n}\sum_{i\neq j\neq k}b_{i j}b_{j k}b_{k i}\right)(n_A)_\mu (n_B)_\nu (n_C)_\sigma \frac{(\hat{r}_{AB})_\theta (\hat{r}_{AC})_\rho (\hat{r}_{BC})_\tau}{\vert r_{AB}\vert^{2\Delta}\vert r_{AC}\vert^{2\Delta}\vert r_{BC}\vert^{2\Delta}}\\
& \left( \text{Tr}\left[ \gamma^\mu \gamma^\theta \gamma^\nu\gamma^\tau\gamma^\sigma\gamma^\rho\right]-\text{Tr}\left[ \gamma^\rho \gamma^\sigma \gamma^\tau\gamma^\nu\gamma^\theta\gamma^\mu\right]
\right)+\dots\, ,
\end{split}
\end{equation}
where the dots denote subleading contributions in the sizes/distances ratios.
Since the trace of a product of gamma matrices does not change if the order of such product is reversed \cite{Peskin:1995ev}, we have just proved that $I_3=0$ at leading order in the long-distance expansion for spin-$1/2$ fields. This is again consistent with the result of \cite{Agon:2021lus}. As argued in the same reference, in the case of a spin-$0$ leading-primary $\mathcal{O}$, the leading-order term in the tripartite information encodes the OPE coefficient $C_{\mathcal{O}\mathcal{O}\mathcal{O}}$. However, in the fermionic case at hand, the fact that the leading coefficient in the tripartite information vanishes identically prevents us from extracting such piece of  CFT data from it.  Hence, in order to start probing the OPE coefficients of the CFT we need to consider either subleading contributions in the long-distance expansions of $I_3$ or $I_2$, or move to the four-partite information, which we discuss in the next section.
\section{Fourpartite information}\label{sec:fourpartite}
Given $4$ disconnected regions $\{A, B, C, D\}$, eq. (\ref{IN_final}) reads
\begin{equation}\label{8pointfc}
\begin{split}
I_4= & \left(R_A R_B R_C R_D\right)^{2\Delta} n_A^{\mu} n_B^{\nu}n_C^{\lambda}n_D^{\eta} \left(\gamma_{\mu}\right)_{\alpha\beta}\left(\gamma_{\nu}\right)_{\rho\sigma}\left(\gamma_{\lambda}\right)_{\pi\zeta}\left(\gamma_{\eta}\right)_{\theta\tau}\mathcal{T}+\dots\, ,
\end{split}
\end{equation}
where we defined
\begin{equation}\label{T}
\mathcal{T}\equiv \lim_{n\rightarrow 1}\frac{1}{n-1}\sum_{i\neq j}\cdots\sum_{r\neq s}b_{ij}b_{k\ell}b_{mn}b_{rs}\left\langle \bar{\psi}_{\alpha}^i (r_A)\psi_{\beta}^j(r_A)\bar{\psi}_{\rho}^k (r_B)\psi_{\sigma}^\ell(r_B)\bar{\psi}_{\pi}^m (r_C)\psi_{\zeta}^n(r_C)\bar{\psi}_{\theta}^r (r_D)\psi_{\tau}^s(r_D)\right\rangle \, .
\end{equation}
As before, we can classify the different contributions to (\ref{T}) according to the number of distinct replica sheets $\mathcal{N}$. Meanwhile, there might be different ways to connect $\mathcal{N}$ vertices of a graph, as long as there are neither sources nor sinks, taking into account permutation of regions and replica copies. For example, for $\mathcal{N}=2$ we have
\begin{figure}[t!]
    \centering
    \begin{subfigure}[t]{0.45\textwidth}
        \centering
        \begin{tikzpicture}
\begin{scope}[thick,decoration={
    markings,
    mark=at position 0.5 with {\arrow{>}}}
    ] 
\draw [postaction={decorate}](-1.5,0) .. controls (-0.5,-1) and (0.5,-1) .. (1.5,0);
\draw [postaction={decorate}](-1.5,0) .. controls (-0.5,0.4) and (0.5,0.4) .. (1.5,0);
\end{scope}
\begin{scope}[thick,decoration={
    markings,
    mark=at position 0.5 with {\arrow{<}}}
    ] 
\draw [postaction={decorate}](-1.5,0) .. controls (-0.5,1) and (0.5,1).. (1.5,0);
\draw [postaction={decorate}](-1.5,0) .. controls (-0.5,-0.4) and (0.5,-0.4).. (1.5,0);
  \end{scope}
\draw [fill] (-1.5,0) circle[radius=0.06]; 
\draw [fill] (1.5,0) circle[radius=0.06];

\node [right] at (1.5,0) {$j$};
\node [left] at (-1.5,0) {$i$};
\node [below] at (0,-0.8) {$A$};
\end{tikzpicture}
        \caption{Graph with $\mathcal{N}=2$. This is responsible for three terms, each corresponding to a different way of labeling the free arrow running from $i$ to $j$.} \label{fig:N2diagram}
    \end{subfigure}
    \hfill
    \begin{subfigure}[t]{0.45\textwidth}
        \centering
        \begin{tikzpicture}
\begin{scope}[ thick,decoration={
    markings,
    mark=at position 0.5 with {\arrow{<}}}
    ] 
\draw [postaction={decorate}](0,0) .. controls (0.707107,0) and (1.41421,0.707107).. (1.41421,1.41421);
\draw [postaction={decorate}](1.41421,1.41421) .. controls (0.707107,1.41421) and (0,0.707107).. (0,0);
\draw [postaction={decorate}](2.82842,0) .. controls (2.12131,0) and (1.41421,0.707107).. (1.41421,1.41421);
\draw [postaction={decorate}](1.41421,1.41421) .. controls (2.12131,1.41421) and (2.82842,0.707107).. (2.82842,0);
\end{scope}
\draw [fill] (0,0) circle[radius=0.06]; 
\draw [fill] (2.82842,0) circle[radius=0.06];
\draw [fill] (1.41421,1.41421) circle[radius=0.06]; 

\node [below] at (0,0) {$i$};
\node [above] at (1.41421,1.41421) {$j$};
\node [below] at (2.82842,0) {$k$};
\node [left] at (0.707107,1.41421) {$A$};
\end{tikzpicture}
\begin{tikzpicture}
\begin{scope}[ thick,decoration={
    markings,
    mark=at position 0.5 with {\arrow{>}}}
    ] 
\draw [postaction={decorate}](2.12132, -0.707105) .. controls (2.12132,0) and (1.41421, 0.707105).. (0.707105, 0.707105);
\draw [postaction={decorate}](0.707105, 0.707105) .. controls (0.707105,0) and (1.41421, -0.707105).. (2.12132, -0.707105);
\draw [postaction={decorate}](2.12132, 2.12132) .. controls (2.12132, 1.41421) and (1.41421, 0.707105).. (0.707105, 0.707105);
\draw [postaction={decorate}](0.707105, 0.707105) .. controls (0.707105, 1.41421) and (1.41421, 2.12132).. (2.12132, 2.12132);
\end{scope}
\draw [fill] (0.707105, 0.707105) circle[radius=0.06]; 
\draw [fill] (2.12132, -0.707105) circle[radius=0.06];
\draw [fill] (2.12132, 2.12132) circle[radius=0.06]; 

\node [left] at (0.707105, 0.707105) {$i$};
\node [right] at (2.12132, -0.707105) {$k$};
\node [right] at (2.12132, 2.12132) {$j$};
\node [above] at (0.707105, 1.41421) {$A$};
\end{tikzpicture}
        \caption{Graph with $\mathcal{N}=3$. Each graph is responsible for six ($3!$) different terms, which correspond to non-equivalent assignations of letters to the arrows} \label{fig:N3diagram}
    \end{subfigure}

    \vspace{1cm}
    \begin{subfigure}[t]{0.45\textwidth}
    \centering
\begin{tikzpicture}
\begin{scope}[ thick,decoration={
    markings,
    mark=at position 0.5 with {\arrow{>}}}
    ] 
\draw [postaction={decorate}] (0,0) --(2,0);
 \draw [postaction={decorate}] (2,0)-- (2,2); 
 \draw [postaction={decorate}] (2,2) --(0,2);
  \draw [postaction={decorate}] (0,2) --(0,0);
  \end{scope}
\draw [fill] (0,0) circle[radius=0.06]; 
\draw [fill] (2,0) circle[radius=0.06];
\draw [fill] (2,2) circle[radius=0.06];
\draw [fill] (0,2) circle[radius=0.06]; 
\node [right] at (2,2) {$k$};
\node [right] at (2,0) {$j$};
\node [left] at (0,0) {$i$};
\node [left] at (0,2) {$\ell$};
\node [below] at (1,0) {$A$};
\end{tikzpicture}
        \caption{Connected graph with $\mathcal{N}=4$. There are $3!$ possible combinations, which are associated to the distinct ways of labeling the free paths.} \label{fig:N4diagram1}
    \end{subfigure}
    \hfill
        \begin{subfigure}[t]{0.45\textwidth}
    \centering
   \begin{tikzpicture}
\begin{scope}[ thick,decoration={
    markings,
    mark=at position 0.5 with {\arrow{>}}}
    ] 
\draw [postaction={decorate}](0,-2) .. controls (0.5,-2.5) and (1.5,-2.5).. (2,-2);
\draw [postaction={decorate}](0,-0.7) .. controls (0.5,-1.2) and (1.5,-1.2).. (2,-0.7);
\end{scope}
\begin{scope}[ thick,decoration={
    markings,
    mark=at position 0.5 with {\arrow{<}}}
    ] 
\draw [postaction={decorate}](0,-2) .. controls (0.5,-1.5) and (1.5,-1.5).. (2,-2);
\draw [postaction={decorate}](0,-0.7) .. controls (0.5,-0.2) and (1.5,-0.2).. (2,-0.7);
  \end{scope}
\draw [fill] (0,-2) circle[radius=0.06]; 
\draw [fill] (2,-2) circle[radius=0.06];
\draw [fill] (0,-0.7) circle[radius=0.06]; 
\draw [fill] (2,-0.7) circle[radius=0.06]; 
\node [left] at (0,-2) {$i$};
\node [right] at (2,-2) {$j$};
\node [left] at (0,-0.7) {$\ell$};
\node [right] at (2,-0.7) {$k$};
\node [below] at (1,-2.5) {$A$};

\end{tikzpicture}
        \caption{Disconnected diagram with $\mathcal{N}=4$. This leads to 3 different terms corresponding to the possible ways of labeling the arrow flowing from $j$ to $i$.} \label{fig:N4diagram2}
    \end{subfigure}
    \captionsetup{width=0.9\textwidth}
    \caption{Diagrams representing the different contributions to $I_4$, classified according to the number of vertices $\mathcal{N}=2$ (\subref{fig:N2diagram}), $\mathcal{N}=3$ (\subref{fig:N3diagram}) and $\mathcal{N}=4$ (\subref{fig:N4diagram1}), (\subref{fig:N4diagram2}).}
\end{figure}
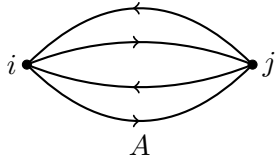
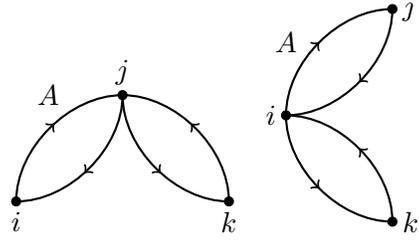
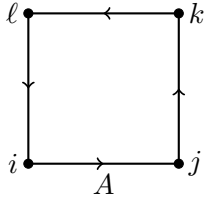
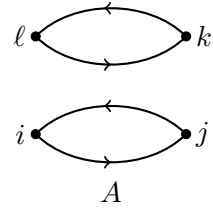
\begin{equation}\label{term1}
\mathcal{T}_{\mathcal{N}=2}=c_1\left[\left\langle \bar{\psi}_{\alpha} (r_A)\bar{\psi}_{\rho} (r_B)\psi_{\zeta}(r_C)\psi_{\tau}(r_D)\right\rangle \left\langle \psi_{\beta}(r_A)\psi_{\sigma}(r_B)\bar{\psi}_{\pi} (r_C)\bar{\psi}_{\theta} (r_D)\right\rangle +(B\leftrightarrow \{C, D\})\right]\, ,
\end{equation}
with
\begin{equation}
c_1\equiv\lim_{n\rightarrow 1} \frac{1}{n-1}\sum_{i\neq j} (b_{ij})^2(b_{ji})^2\, .
\end{equation}
In the first term correlators feature $A$ and $B$ spinors in the same representation, so that corresponds to the graph of Fig.\,\ref{fig:N2diagram} in which these regions flow in the same direction. Furthermore, there are two additional contributions, which we get by permuting the label $B$ with $C$ and $D$. These terms have the same overall sign because bilinears $\bar{\psi}\psi$ commute. 

For $\mathcal{N}=3$ we find\footnote{Lorentz indices are implicit to avoid cluttering, but each spinor is labelled according to (\ref{8pointfc}). For example, $\psi_A \equiv \psi^j_\alpha (r_A)$, $\bar{\psi}_D\equiv\bar{\psi}_\theta^r(r_D)$ and so on. }
\begin{equation}\label{term2}
\begin{split}
\mathcal{T}_{\mathcal{N}=3}= c_2&\left[ \left\langle\psi_A\bar{\psi}_B\psi_C\bar{\psi}_D \right\rangle\left\langle\bar{\psi}_A \psi_B\right\rangle\left\langle \bar{\psi}_C\psi_D\right\rangle + \left\langle\bar{\psi}_A\psi_B\bar{\psi}_C\psi_D \right\rangle \left\langle\psi_A \bar{\psi}_B\right\rangle\left\langle \psi_C\bar{\psi}_D\right\rangle \right. \\ &\left.+ \text{permutations of}~ \{B,C,D\}\right]\, ,
\end{split}
\end{equation}
with 
\begin{equation}
c_2\equiv\lim_{n\rightarrow 1} \frac{1}{n-1}\sum_{i\neq j\neq k} b_{ij}b_{ji}b_{k i}b_{i k}\, .
\end{equation}
The first two terms correspond to a particular labeling of the free arrows in Fig.\,\ref{fig:N3diagram}. Permutation of regions $\{B, C, D\}$ leads to $2\times 3!$ additional terms. Note that the right graph in Fig.\,\ref{fig:N3diagram} is equivalent to reversing the direction of the arrow assigned to region $A$.

On the other hand, we separate the $\mathcal{N}=4$ case into a connected piece 
\begin{equation}\label{square}
\mathcal{T}_{\mathcal{N}=4}^{\text{conn}}=c_3\left[\left\langle \bar{\psi}_A\psi_D\right\rangle \left\langle \psi_A \bar{\psi}_B\right\rangle \left\langle \psi_B\bar{\psi}_C \right\rangle \left\langle \psi_C \bar{\psi}_D\right\rangle +\text{permutations of}~\{ B,C,D\}\right]
\end{equation}
where
\begin{equation}
c_3\equiv\lim_{n\rightarrow 1} \frac{1}{n-1}\sum_{i\neq j\neq k\neq \ell} b_{ij}b_{k\ell}b_{\ell i}b_{j k}\, ,
\end{equation}
with $3!$ different contributions corresponding to the possible ways to label the $3$ free sides of the square (see Fig.\,\ref{fig:N4diagram1}), and a disconnected piece (see Fig.\,\ref{fig:N4diagram2})
\begin{equation}\label{term4}
\mathcal{T}_{\mathcal{N}=4}^{\text{disconn}}=c_4\left[ \left\langle \bar{\psi}_A\psi_B\right\rangle \left\langle \psi_A \bar{\psi}_B\right\rangle \left\langle \bar{\psi}_C\psi_D \right\rangle \left\langle\psi_C \bar{\psi}_D\right\rangle + (B\leftrightarrow \{C,D\})\right]\, ,
\end{equation}
\begin{equation}
c_4\equiv\lim_{n\rightarrow 1} \frac{1}{n-1}\sum_{i\neq j\neq k\neq \ell} b_{ij}b_{ji}b_{\ell k}b_{k\ell}\, .
\end{equation}
Finally, we have 
\begin{equation}
\mathcal{T}=\mathcal{T}_{\mathcal{N}=2}+\mathcal{T}_{\mathcal{N}=3}+\mathcal{T}_{\mathcal{N}=4}^{\text{conn}}+\mathcal{T}_{\mathcal{N}=4}^{\text{disconn}}
\end{equation}
which can be schematically arranged as
\begin{equation}\label{I4schematic2}
\begin{split}\mathcal{T}=&c_1 \left\langle \psi^4\right\rangle_{\text{conn}}^2+(c_2+c_1)  \left\langle \psi^4\right\rangle_{\text{conn}} \left\langle \psi^2\right\rangle^2+(c_3+2 c_2+c_1) \left\langle \psi^2\right\rangle^4\\&+(c_4+4 c_2+2 c_1)\left\langle \psi^2\right\rangle^2 \left\langle \psi^2\right\rangle^2\, ,
\end{split}
\end{equation}
in terms of the connected part of the four point function $\left\langle \psi^4\right\rangle_{\text{conn}}$
\begin{equation}
\left\langle \bar{\psi}_A\bar{\psi}_B\psi_C\psi_D\right\rangle_{\text{conn}}\equiv \left\langle \bar{\psi}_A\bar{\psi}_B\psi_C\psi_D\right\rangle+ \left\langle \bar{\psi}_A\psi_C\right\rangle \left\langle \bar{\psi}_B\psi_D\right\rangle-\left\langle \bar{\psi}_A\psi_D\right\rangle \left\langle \bar{\psi}_B\psi_C\right\rangle\, .
\end{equation}
The reason for organizing four-partite information in this way is that it simplifies manifestly for Gaussian models (either free local theories or holographic), because in those scenarios the connected correlators vanish. 

Since the coefficient (\ref{b_final}) is almost identical to the one for a CFT with scalar lightest primary, except only for the $\text{sgn}(i-j)$ function responsible for the antisymmetry of $b_{ij}$, we just comment on the slight differences with the scalar case regarding the computation of the coefficients involved in (\ref{I4schematic2}) and refer the reader to \cite{Agon:2022efa} for a detailed derivation.  On the one hand, it is easy to write $c_1$, $c_2$ and $c_3$ in terms of the coefficients $c_{4:2}^{(4)}$, $c_{4:3}^{(2,2)}$ and $c_{4:4}^{(2,2)}$ appearing in \cite{Agon:2022efa},
\begin{eqnarray}\label{c1def}
c_1 & = & 2 c_{4:2}^{(4)}\,\,\,=+\frac{\Gamma^2(4\Delta +1)}{\Gamma(8\Delta +2)}2^{8\Delta}\, ,\\
c_2 & = & 2 c_{4:3}^{(2,2)}=-\frac{\Gamma^2(4\Delta +1)}{\Gamma(8\Delta +2)}2^{8\Delta}\, ,\\
c_4 & = & 8 c_{4:4}^{(2,2)} =+\frac{\Gamma^2(4\Delta +1)}{\Gamma(8\Delta +2)}2^{8\Delta+1}\, .
\end{eqnarray}
Meanwhile, 
\begin{equation}
c_3=\lim_{ n\to 1}\frac{2 n}{(n-1)}\[
\sum_{l=3}^{n-1}\sum_{k=2}^{l-1}\sum_{j=1}^{k-1}+
\sum_{k=3}^{n-1}\sum_{l=2}^{k-1}\sum_{j=1}^{l-1}+
\sum_{l=3}^{n-1}\sum_{j=2}^{l-1}\sum_{k=1}^{j-1}\]b_{0j}b_{jk}b_{kl}b_{l0}\, ,
\end{equation}
which is related to the scalar coefficients $C_{ij}$ through
\begin{equation}
c_3=\lim_{ n\to 1}\frac{2 n}{(n-1)}\[-
\sum_{l=3}^{n-1}\sum_{k=2}^{l-1}\sum_{j=1}^{k-1}+
\sum_{k=3}^{n-1}\sum_{l=2}^{k-1}\sum_{j=1}^{l-1}+
\sum_{l=3}^{n-1}\sum_{j=2}^{l-1}\sum_{k=1}^{j-1}\]C_{0j}C_{jk}C_{kl}C_{l0}\, .
\end{equation}
Note that there is a relative minus sign changing the first term, precisely due to the presence of the sgn function. This prevents $c_3$ from being proportional to $c_{4:4}^{(1,1,1,1)}$. Following the conventions of \cite{Agon:2022efa}, we have instead
\begin{equation}\label{c3def}
c_3=8(J_1- J_2 +J_3)\, ,
\end{equation}
where
\begin{eqnarray}\label{J1}
J_1(\Delta)&\equiv &\frac{2^{8\Delta-8}}{\pi^6}\int_{-\infty}^{\infty} \mathrm{d}p \int_{-\infty}^{\infty} \mathrm{d}q\int_{-\infty}^{\infty} \mathrm{d}r B_p(\Delta)B_q(\Delta) B_r(\Delta)\frac{2 B_{p+q+r}(\Delta+1)}{(e^{p+r}-1)(e^{p+q}-1)}\, ,\\\label{J2}
J_2(\Delta)&\equiv&\frac{2^{8\Delta-8}}{\pi^6}\int_{-\infty}^{\infty} \mathrm{d}p \int_{-\infty}^{\infty} \mathrm{d}q\int_{-\infty}^{\infty} \mathrm{d}r B_p(\Delta)B_q(\Delta) B_r(\Delta)\frac{3 e^{\frac{q+r}{2}}B_{p}(\Delta+1)}{(e^p-e^r)(e^p-e^q)}\, ,\\\label{J3}J_3(\Delta)&\equiv&\frac{2^{8\Delta-8}}{\pi^6}\int_{-\infty}^{\infty} \mathrm{d}p \int_{-\infty}^{\infty} \mathrm{d}q\int_{-\infty}^{\infty} \mathrm{d}r B_p(\Delta)B_q(\Delta) B_r(\Delta)\frac{4 e^{\frac{p+r}{2}} B_{q}(\Delta+1)}{(e^{p+r}-1)(e^{q+r}-1)}\, ,
\end{eqnarray}
and
\begin{equation}
B_q(\Delta)\equiv \frac{\vert \Gamma(\Delta+ i \frac{q}{2\pi})\vert^2}{\Gamma(2\Delta)}\, .
\end{equation}
An important observation is that $c_4=-4 c_2 -2 c_1$, so in agreement with the clustering principle 
the last term in (\ref{I4schematic2}) vanishes.\footnote{By clustering we mean the fact that any measure of $N-$party spatial correlations should vanish when any subset of the parties is separated from the rest by an infinite distance. For the $N-$partite information, this follows from the fact that the entropy of two subsets of regions equals the sum of their entropies when they are at infinite separation, together with definition (\ref{IN_definition}).} The second term also disappears because $c_2=-c_1$. The final general expression for the four-partite information leading-contribution is then given by (\ref{8pointfc}), where
\begin{align}\label{I4schematic22}
\mathcal{T}=&+c_1 \left[\left\langle \bar{\psi}_A\bar{\psi}_B \psi_C\psi_D\right\rangle_{\rm conn} \left\langle \psi_A\psi_B\bar{\psi}_C\bar{\psi}_D\right\rangle_{\rm conn} +(B\leftrightarrow \{C, D\})\right]
\\ \notag &+(c_3-c_1) \left[\left\langle \bar{\psi}_A\psi_D\right\rangle \left\langle \psi_A \bar{\psi}_B\right\rangle \left\langle \psi_B\bar{\psi}_C \right\rangle \left\langle \psi_C \bar{\psi}_D\right\rangle +\text{permutations of}~\{ B,C,D\}\right]\, .
\end{align}

\subsection{Free-fermion CFT}
When the fermion is free only the second term in (\ref{I4schematic22}) contributes. We call its coefficient \begin{equation}\label{c_free_d}
c_{\text{free}}(d)\equiv c_3\left(\frac{d-1}{2}\right)-c_1\left(\frac{d-1}{2}\right)\, ,
\end{equation}
where we used the corresponding free-fermion conformal dimension
\begin{equation}
\Delta_{\rm free\, fermion}=\frac{d-1}{2}\, .
\end{equation}
If we assume that the regions are all situated at $t=0$, we find explicitly, 
\begin{equation}\label{I4_free}
\begin{split}
 I_4 &= c_{\text{free}}(d) 2^{[(d+2)/2]} (R_A R_B R_C R_D)^{d-1}\times  \\
\Bigg\{ &+ \frac{\left[ \left(\hat{r}_{AB}\cdot \hat{r}_{AD}\right)\left(\hat{r}_{BC}\cdot \hat{r}_{CD}\right)+\left(\hat{r}_{AB}\cdot \hat{r}_{BC}\right)\left(\hat{r}_{AD}\cdot \hat{r}_{CD}\right)-\left(\hat{r}_{AB}\cdot \hat{r}_{CD}\right)\left(\hat{r}_{AD}\cdot \hat{r}_{BC}\right)\right]}{ (\vert r_{AB}\vert\vert r_{AD}\vert\vert r_{BC}\vert\vert r_{CD}\vert)^{d-1} }\\
&-  \frac{\left[ \left(\hat{r}_{AB}\cdot \hat{r}_{AC}\right)\left(\hat{r}_{BD}\cdot \hat{r}_{CD}\right)+\left(\hat{r}_{AB}\cdot \hat{r}_{BD}\right)\left(\hat{r}_{AC}\cdot \hat{r}_{CD}\right)-\left(\hat{r}_{AB}\cdot \hat{r}_{CD}\right)\left(\hat{r}_{AC}\cdot \hat{r}_{BD}\right)\right]}{(\vert r_{AB}\vert\vert r_{AC}\vert\vert r_{BD}\vert\vert r_{CD}\vert)^{d-1}}\\
&- \frac{\left[ \left(\hat{r}_{AC}\cdot \hat{r}_{AD}\right)\left(\hat{r}_{BC}\cdot \hat{r}_{BD}\right)+\left(\hat{r}_{AC}\cdot \hat{r}_{BC}\right)\left(\hat{r}_{AD}\cdot \hat{r}_{BD}\right)-\left(\hat{r}_{AC}\cdot \hat{r}_{BD}\right)\left(\hat{r}_{AD}\cdot \hat{r}_{BC}\right)\right]}{(\vert r_{AC}\vert\vert r_{AD}\vert\vert r_{BC}\vert\vert r_{BD}\vert)^{d-1}} \Bigg\}\, .
\end{split}
\end{equation}
Moreover, if $d=3$ the integral expressions in (\ref{J1}), (\ref{J2}) and (\ref{J3}) evaluate to
\bea
J_1(1)=\frac{64}{945}-\frac{4}{27 \pi^2}\, \quad J_2(1)=\frac{32}{315}+\frac{4}{9 \pi^2}\quad {\rm and}\quad J_3(1)=\frac{32}{945}-\frac{8}{27 \pi^2}
\eea
respectively. Hence, 
\bea
c_3(\Delta=1)=-\frac{64}{9\pi^2} \quad \rightarrow \quad c_{\rm free}(d=3)=-\frac{64}{9\pi^2}-\frac{128}{315}\approx  -1.12686\, .
\eea
In order to test (\ref{I4_free}) numerically, we focus on two particularly simple arrangements, namely one in which the regions are located at the vertices of a square of length $r$, and another one in which the regions are collinear and separated a distance $r$. For the free fermion we get,
\begin{align}\label{sq}
I_4^{\text{square}} &= -2 \(\frac{64}{9\pi^2}+\frac{128}{315} \)\frac{R^8}{r^8}+ \text{subleading}\, ,\\  \label{sq1}
I_4^{\text{colinear}} & =-\frac{1}{6}\(\frac{64}{9\pi^2}+\frac{128}{315} \)  \frac{R^8}{r^8}+ \text{subleading}\, . \end{align}
We test these exact predictions in the lattice in the following subsection.

\subsubsection{Lattice calculations}
In order to test our new formula for the four-partite information, we perform numerical calculations in the lattice for a free spin-$1/2$ field in $d=3$. For that, we consider 
 fields  $\psi_i$ defined at each lattice point $i$ and satisfying canonical   anticommutation  relations $\{\psi_i,  \psi_j^{\dagger} \}=\delta_{ij}$. Given some Gaussian state  $\rho$, we define the matrix of correlators  $D_{ij} \equiv \rm{Tr} ( \rho \psi_i \psi_j^{\dagger}  )$. Then, the entanglement entropy of a given entangling region $A$ can be computed  as \cite{Casini:2009sr}
 \begin{equation}\label{eeG}
S (A)= - {\rm Tr}   \left[  D_A   \log  D_A   + (1- D_A) \log (1- D_A)\right] \, ,
\end{equation}
where $D_A $ is the  restriction of the correlators matrix to the  lattice sites inside $A$. From this it is straightforward to compute any $N$-partite information by considering linear combinations of entanglement entropies of individual regions and their unions. The Hamiltonian for the free fermion in the lattice can be written as
\begin{equation}
H=-\frac{i}{2}  \sum_{n ,m} \left[  \left(\psi^{\dagger}_{m, n}   \gamma^0     \gamma^1 (\psi_{m+1,n}  -\psi_{m,n})+\psi^{\dagger}_{m,  n}   \gamma^0 \gamma^2   (\psi_{m,n+1}   -\psi_{ m,n}  ) \right)  - {\rm h.c.}\right] \, ,
\end{equation}
and the vacuum state correlators, which are the ones of interest here, read \cite{Casini:2009sr} 
\begin{equation}
D_{(n,k), (j,l)} =   \frac{1}{2} \delta_{n,  j} \delta_{kl}     - \int_{-\pi}^{\pi} \mathrm{d} x  \int_{-\pi}^{\pi } \mathrm{d} y   \frac{\sin (x)   \gamma^0  \gamma^1+   \sin(y) \gamma^0  \gamma^2}{8\pi^2 \sqrt{   \sin^2   x +   \sin^2   y}}  e^{i(x (n-j)  +y(k-l))}\, .
\end{equation}
A systematic evaluation of these correlators combined with Eq.\,(\ref{eeG}) allows us to compute the vacuum four-partite information of the free fermion for arbitrary lattice regions.

\begin{figure}[t]
\begin{center}  
\includegraphics[width=0.35\textwidth]{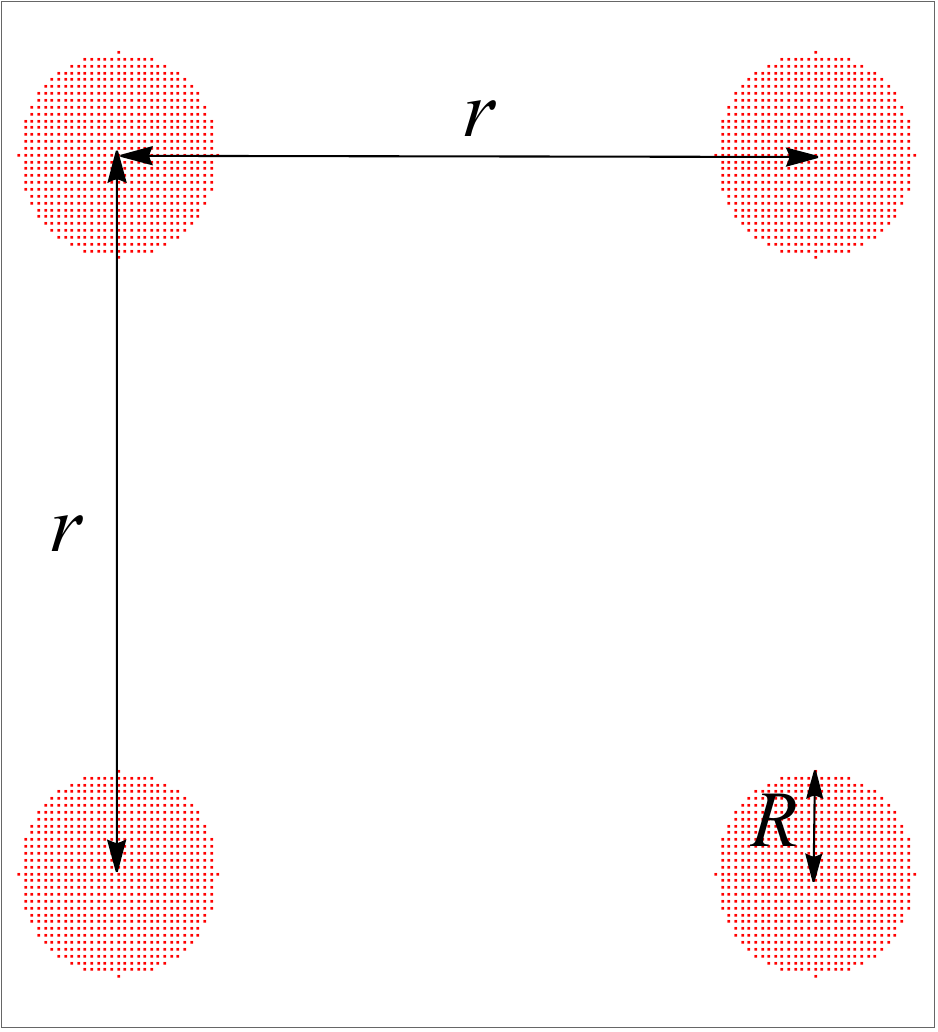}\\ \vspace{0.3cm}
\includegraphics[width=0.83\textwidth]
{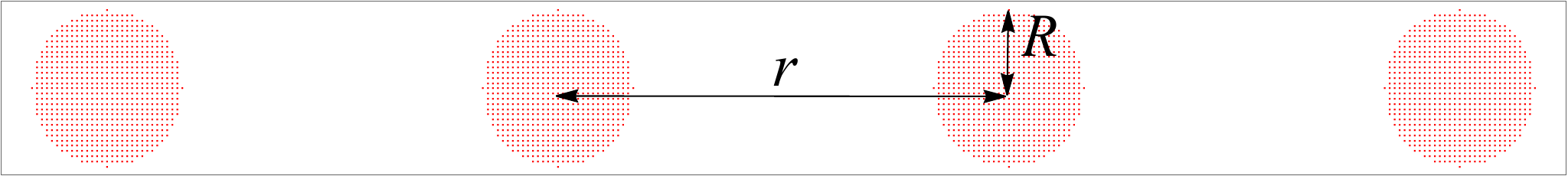}
\captionsetup{width=0.9\textwidth}
\caption{We show the two geometric arrangements of disk regions considered for the evaluation of the $I_4$ of a free fermion in the lattice. }
\label{I44}
\end{center}  
\end{figure}

In the continuum limit, the results converge to the ones of the actual CFT of a free Dirac field.
In order to extract those continuum-theory values from our lattice calculations, we proceed as follows. Given a set of four entangling regions and a particular geometric arrangement of those, we compute $I_4$ for increasingly greater values of the characteristic size of the regions, while keeping fixed the proportions of the arrangement. In particular, we consider four disk regions of radius $R$ and two geometric arrangements. As shown in Fig.\,\ref{I44}, we locate their centers at the vertices of a square of side length $r$ as well as in a colinear distribution with consecutive centers separated a distance $r$. In each case, keeping $R/r\equiv x$ fixed while varying $R$ and $r$ produces a collection of values of $I_4$. Those can be fitted using  functions $\{1,1/x,1/x^2,\dots\}$ in order to extract the constant value in the $R,r\rightarrow \infty$ limit. Reliable limiting values do not depend on  the order of the last fitting function. One also needs to take into account the ``doubling'' in the number of fermionic degrees of freedom which takes place in the lattice. In $d=3$, the Dirac fermion result is obtained by dividing the final result by 4.

\begin{figure}[t]
\begin{center}  
\includegraphics[width=0.495\textwidth]{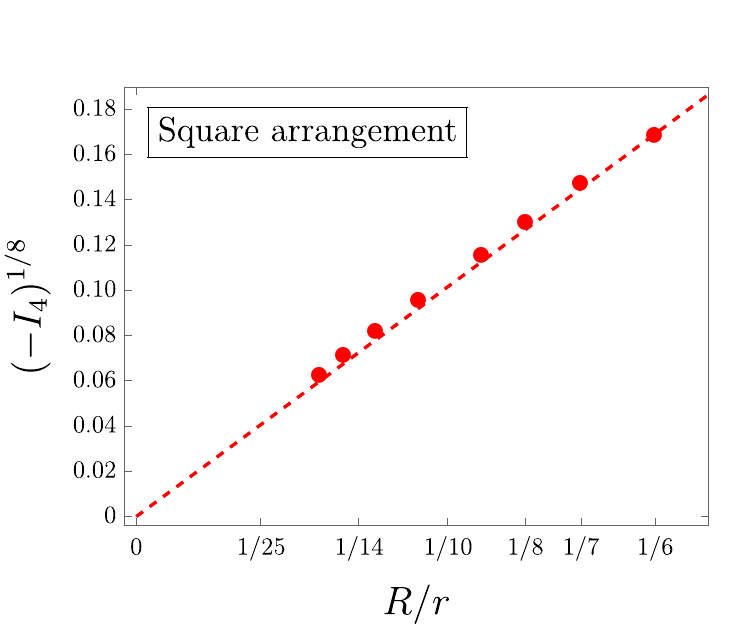} \hspace{-0.2cm}
\includegraphics[width=0.495\textwidth]
{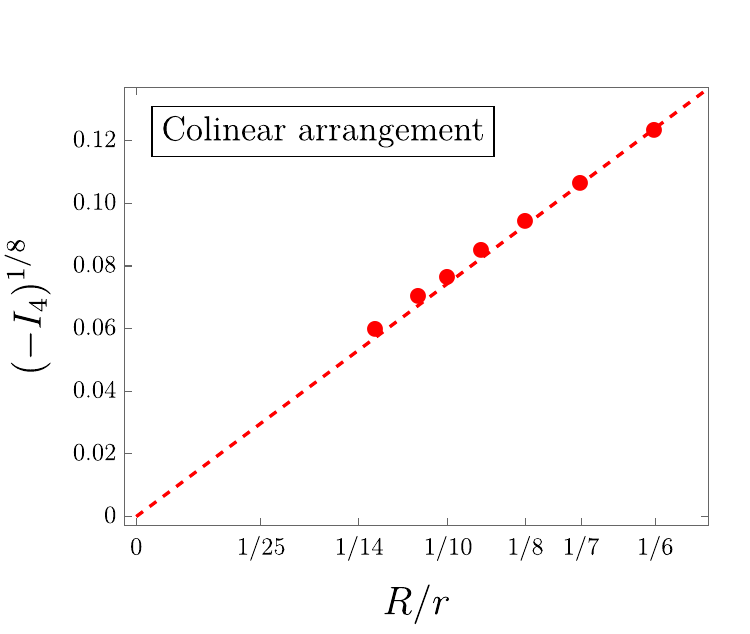}
\captionsetup{width=0.9\textwidth}
\caption{We plot the four-partite information (per degree of freedom) for a three-dimensional free fermion in the lattice. In both plots the four entangling regions correspond to disks of radius $R$. In the left plot they are arranged so that the centers of the disks form a square of length $r$. In the right plot they are arranged so that the centers of the disk all lie within the same straight line, separated consecutively by a distance $r$. The dashed red lines correspond to the theoretical predictions obtained in the main text for the leading term in the $R/r\ll 1$ regime. }
\label{I4r}
\end{center}  
\end{figure} 

The data points obtained from this procedure for the two arrangements are displayed in Fig.\,\ref{I4r}. As we can see, in both cases the theoretical predictions of Eqs.\,(\ref{sq}) and (\ref{sq1}) fit the data very well. The sign, the power of $(R/r)$ and the magnitude of the coefficient are all correctly reproduced by the lattice results. Notice the negative sign of $I_4$ for the free fermion for both arrangements. Then, it is natural to speculate that $I_4<0$ for all possible sets of regions and arrangements. We prove this conjecture to be wrong in the next subsection. There, we rewrite (\ref{I4_free}) in terms of a minimal set of geometrical parameters which completely characterize the arrangement of four spheres in the plane and show that configurations for which $I_4>0$ and $I_4=0$ also exist.
\subsubsection{Geometry of the $I_4$ for $d=3$}
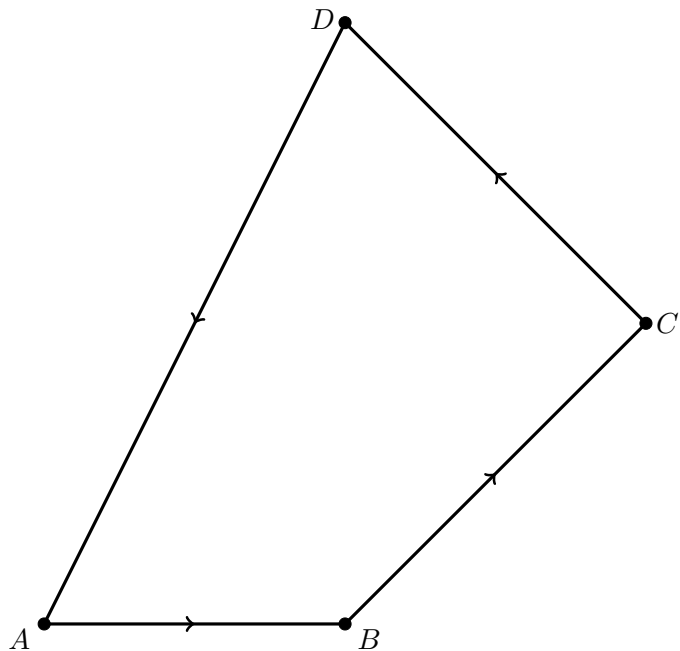
\begin{figure}[h!]
\centering
\begin{tikzpicture}
\begin{scope}[very thick,decoration={
    markings,
    mark=at position 0.5 with {\arrow{>}}}
    ] 
\draw [postaction={decorate}] (0,0) --(4,0);
 \draw [postaction={decorate}] (4,0) --(8,4);
 \draw [postaction={decorate}] (8,4) --(4,8);
  \draw [postaction={decorate}] (4,8) --(0,0);
  \end{scope}
\draw [fill] (0,0) circle[radius=0.08]; 
\draw [fill] (4,0) circle[radius=0.08];
\draw [fill] (8,4) circle[radius=0.08];
\draw [fill] (4,8) circle[radius=0.08]; 
\node [below, left] at (-0.05,-0.2) {$A$};
\node [below, right] at (4.03,-0.2) {$B$};
\node [right] at (8,4) {$C$};
\node [above, left] at (4,8.05) {$D$};
\end{tikzpicture}
\caption{Graph representation of a configuration of four spheres.}
\label{conf-spheres}
\end{figure}

\begin{figure}[h!]
\centering
\begin{tikzpicture}
\begin{scope}[very thick,decoration={
    markings,
    mark=at position 0.5 with {\arrow{>}}}
    ] 
 \draw [postaction={decorate}] (0,0) --(1.68,0);
\draw [postaction={decorate}] (0,0) --(1.19,1.19);
 \draw [postaction={decorate}] (0,0) --(-1.19,1.19);
  \draw [postaction={decorate}] (0,0) --(-0.75,-1.5);
  \end{scope}
\draw [fill] (0,0) circle[radius=0.06]; 
\draw [fill] (1.68,0) circle[radius=0.06];
\draw [fill] (1.19,1.19) circle[radius=0.06];
\draw [fill] (-1.19,1.19) circle[radius=0.06]; 
\draw [fill] (-0.75,-1.5) circle[radius=0.06]; 
\node [below, right] at (1.68,0)  {$\hat{r}_{AB}$};
\node [above, right] at (1.19,1.3) {$\hat{r}_{BC}$};
\node [above, left] at (-1.0,1.5) {$\hat{r}_{CD}$};
\node [below, left] at (-0.7,-1.6) {$\hat{r}_{DA}$};

\node  at (0.9,0.3)  {$\theta_1$};
\node  at (0,0.7) {$\theta_2$};
\node at (-0.6,-0.1) {$\theta_3$};
\node  at (0.4,-0.5) {$\theta_4$};
\end{tikzpicture}
\begin{tikzpicture}
\begin{scope}[very thick,decoration={
    markings,
    mark=at position 0.5 with {\arrow{>}}}
    ] 
 \draw [postaction={decorate}] (0,0) --(1.68,0);
\draw [postaction={decorate}] (0,0) --(1.5,0.75);
 \draw [postaction={decorate}] (0,0) --(-1.19,1.19);
  \draw [postaction={decorate}] (0,0) --(0,-1.68);
  \end{scope}
\draw [fill] (0,0) circle[radius=0.06]; 
\draw [fill] (1.68,0) circle[radius=0.06];
\draw [fill] (1.5,0.75) circle[radius=0.06];
\draw [fill] (-1.19,1.19) circle[radius=0.06]; 
\draw [fill] (0,-1.68) circle[radius=0.06]; 
\node [below, right] at (1.68,0)  {$\hat{r}_{AB}$};
\node [above, right] at (1.5,0.75) {$\hat{r}_{AC}$};
\node [above, left] at (-1.0,1.5) {$\hat{r}_{CD}$};
\node [below, left] at (-0,-1.68) {$\hat{r}_{DB}$};

\node  at (1.1,0.25)  {$\theta_1'$};
\node  at (0.2,0.7) {$\theta'_2$};
\node at (-0.5,-0.2) {$\theta'_3$};
\node  at (0.5,-0.5) {$\theta'_4$};
\end{tikzpicture}
\begin{tikzpicture}
\begin{scope}[very thick,decoration={
    markings,
    mark=at position 0.5 with {\arrow{>}}}
    ] 
\draw [postaction={decorate}] (0,0) --(1.5,0.75);
\draw [postaction={decorate}] (0,0) --(1.19,1.19);
  \draw [postaction={decorate}] (0,0) --(-0.75,-1.5);
\draw [postaction={decorate}] (0,0) --(0,-1.68);
  \end{scope}
\draw [fill] (0,0) circle[radius=0.06]; 
\draw [fill] (1.5,0.75) circle[radius=0.06];
\draw [fill] (1.19,1.19) circle[radius=0.06]; 
\draw [fill] (-0.75,-1.5) circle[radius=0.06]; 
\draw [fill] (0,-1.68) 
circle[radius=0.06];

\node [above, right] at (1.5,0.75) {$\hat{r}_{AC}$};
\node [above, right] at (1.19,1.3) {$\hat{r}_{BC}$};
\node [below, left] at (-0.7,-1.6) {$\hat{r}_{DA}$};
\node [below, left] at (0.2,-1.9) {$\hat{r}_{DB}$};
\node  at (1.25,0.9)  {$\theta''_1$};
\node  at (-0.4,0.3) {$\theta''_2$};
\node at (-0.3,-1.3) {$\theta''_3$};
\node  at (0.6,-0.3) {$\theta''_4$};
\end{tikzpicture}
\caption{Graph representation of the various unit vectors associated with regions $A$, $B$, $C$ and $D$ for the geometric configuration illustrated in Fig.\,\ref{conf-spheres}.}
\label{unit-vecs}
\end{figure}
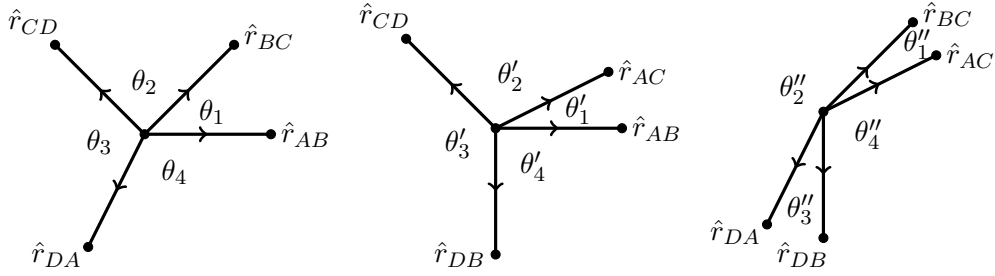

We analyze the result for the $I_4$ in (\ref{I4_free}) with $d=3$ from a geometric standpoint. We start by considering a generic arrangement of the four spheres, as in Fig.\,\ref{conf-spheres}. In Fig.\,\ref{unit-vecs}, we arrange the various unit vectors that describe the relative locations of the spheres in three diagrams, each having relative angles between the unit vectors that add up to $2\pi$.  

These angles satisfy the additional relations  
\bea
\theta_1+\theta_2&=&\theta'_1+\theta'_2\,, \qquad \theta_2+\theta_3=\theta''_2\,, \\
\theta_3+\theta_4&=&\theta'_3+\theta'_4\,, \qquad \theta_1+\theta_4=\theta''_1+\theta''_3+\theta''_4\,,
\eea
as well as 
\bea\label{theta13}
\theta_1-\theta'_1=\theta''_1\,, \qquad {\rm and}\qquad  \theta'_3-\theta_3=\theta''_3\,.
\eea
In terms of them, the $I_4$ in (\ref{I4_free}) takes the form
\begin{equation}\label{I4_free-geo}
\begin{split}
 I_4 &= 4\,c_{\text{free}}(3)(R_A R_B R_C R_D)^{2}\times  \\
\Bigg\{ &- \frac{ \cos\(\theta_1+\theta_3\)}{ (\vert r_{AB}\vert\vert r_{AD}\vert\vert r_{BC}\vert\vert r_{CD}\vert)^{2} }+  \frac{ \cos\(\theta'_1+\theta'_3\)}{(\vert r_{AB}\vert\vert r_{AC}\vert\vert r_{BD}\vert\vert r_{CD}\vert)^{2}}- \frac{ \cos\(\theta''_1-\theta''_3\)}{(\vert r_{AC}\vert\vert r_{AD}\vert\vert r_{BC}\vert\vert r_{BD}\vert)^{2}} \Bigg\}\, .
\end{split}
\end{equation}
Using relations (\ref{theta13}) one can write the orientation dependence of the $I_4$ in terms of the two angles $\theta_1+\theta_3$ and $\theta'_1+\theta'_3$. Namely, the last term in (\ref{I4_free-geo}) can be written as 
\bea
\cos\(\theta''_1-\theta''_3\)=\cos\[(\theta_1+\theta_3)-(\theta'_1+\theta'_3)\]\,.
\eea
Generically the signs of each term in the above formula depends on the specific geometric arrangement of the spheres. However, the distances that appear in the above formula depend also on those angles and thus it is possible that upon a closer scrutiny the overall sign of an arbitrary configuration is always the same. We explore this question next. 

We take as our independent variables the distances $r_{AB}$, $r_{BC}$ and $r_{CD}$ and the angles $\theta_1$ and $\theta_2$. The remaining distances can be written explicitly in terms of these variables as 
\bea\label{rels-distances}
r_{AC}^2&=&r^2_{AB}+r^2_{BC}+2\,r_{AB} r_{BC}\cos\theta_1\,, \nonumber\\
r_{BD}^2&=&r^2_{BC}+r^2_{CD}+2\,r_{BC} r_{CD}\cos\theta_2\,,\\
r_{AD}^2&=&r^2_{AB}+r^2_{BC}+r^2_{CD}+2\,r_{AB} r_{BC}\cos\theta_1+2\,r_{BC} r_{CD}\cos\theta_2+2\,r_{AB} r_{CD}\cos(\theta_1+\theta_2)\,. \nonumber
\eea
Similarly, using the following simple relations:
\bea
r_{AC}\sin\theta'_1&=&r_{BC}\sin\theta_1 \,,\nonumber\\
r_{BD}\sin\(\theta_1 +\theta_2+\theta_3'-\pi\)&=&r_{BC}\sin\theta_1+r_{CD}\sin\(\theta_1+\theta_2\) \,,\nonumber\\
r_{AD}\sin\(\theta_1 +\theta_3+\theta_2-\pi\)&=&r_{BC}\sin\theta_1+r_{CD}\sin\(\theta_1+\theta_2\) \,,
\eea
together with:
\bea
r_{AC}\cos\theta'_1&=&r_{AB}+r_{BC}\cos\theta_1 \,,\nonumber\\
r_{BD}\cos\(\theta_1 +\theta_2+\theta'_3\)&=&-r_{BC}\cos\theta_1-r_{CD}\cos\(\theta_1+\theta_2\) \,,\nonumber\\
r_{AD}\cos\(\theta_1 +\theta_2+\theta_3\)&=&-r_{AB}-r_{BC}\cos \theta_1-r_{CD}\cos\(\theta_1+\theta_2\) \,,
\eea
we can write formulas for the angles of interest $\theta_1+\theta_3$, $\theta'_1$ and $\theta'_3$ as a function of the independent parameters. 
Next we rewrite $I_4$ as
\bea
I_4 = 4\,c_{\text{free}}(3)\frac{(R_A R_B R_C R_D)^{2}}{r^8}g[b,c;\theta_1, \theta_2]\
\eea
where the function
\bea \label{g}
g[b,c;\theta_1, \theta_2]=- \frac{b^2 c^2\, \cos\(\theta_1+\theta_3\)}{ \tilde{r}^2_{AD} }+  \frac{c^2\, \cos\(\theta'_1+\theta'_3\)}{\tilde{r}^2_{AC}\tilde{r}^2_{BD}}- \frac{ b^2\cos\[(\theta_1+\theta_3)-(\theta'_1+\theta'_3)\]}{\tilde{r}^2_{AC}\tilde{r}^2_{AD}\tilde{r}^2_{BD}} 
\eea
determines the sign of $I_4$.
In the above formula, we rescaled our distance parameters 
$r_{AB}=r$, $r_{BC}=r/b$, and $r_{CD}=r/c$ with $\{b,c\}\in[0,1]$ and defined the rescaled dependent distances (\ref{rels-distances}) as $\tilde{r}_{AC}\equiv r_{AC}/r_{AB}$, $\tilde{r}_{AD}\equiv r_{AD}/r_{AB}$ and $\tilde{r}_{BD}\equiv r_{BD}/r_{AB}$. Thus, $g[b,c;\theta_1, \theta_2]$ depends on a four-dimensional compact space given by $\{b,c\}\in[0,1]$ and the angles $\{\theta_1, \theta_2\}\in [0,\pi]$. 

\begin{figure}[t!]
\begin{center}
\includegraphics[width=0.5\textwidth]{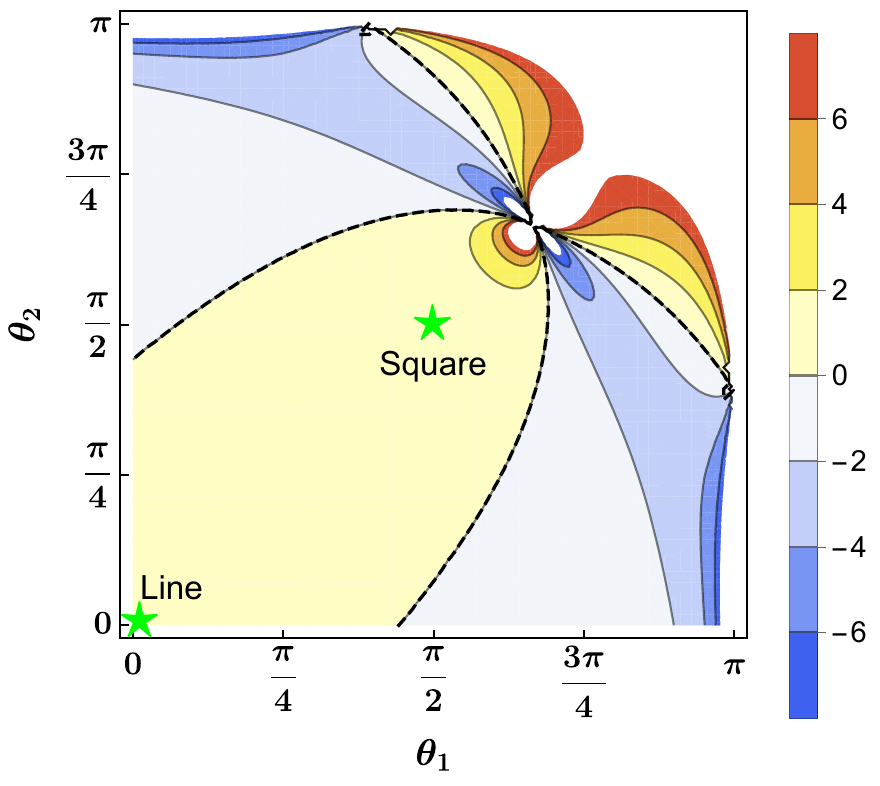}     \captionsetup{width=0.9\textwidth}
\caption{Contour plot for $g[b=1,c=1; \theta_1,\theta_2]$. Dashed black lines represent configurations with vanishing leading term for four-partite information at large distances. Green stars mark the special cases of spheres placed in the vertices of a square or within a line, corresponding to $(\theta_1,\theta_2)=(\pi/2,\pi/2)$, and $(\theta_1,\theta_2)=(0,0)$, respectively.}
\label{fig:g_b1_c1}
\end{center}
\end{figure}

In Fig.\,\ref{fig:g_b1_c1} we numerically plot (\ref{g}) for $b=c=1$. There, we show that $I_4$ is not always negative, as happens with the square and colinear arrangments tested in the lattice, but there is also a continuous set of geometrical configurations for which it turns out to be either positive or zero. 

\subsection{Non-free CFT}
Let us now say a few more things about the case of a general CFT. In that case we can express the four-point function as a linear combination of tensor structures, each coefficient being an arbitrary function of the cross ratios determined by the dynamics of the theory. In order to deduce such expansion, we resort to the embedding space formalism, as explained in \cite{Isono_2017}.

Based on the homogeneity property of the embedding space spinors, as well as the Lorentz invariance of the scalar correlators, we know that
\begin{equation}\label{4point_general}
\left\langle \Psi (X_1,\bar{S_1})\bar{\Psi}(X_2,S_2)\Psi (X_3,\bar{S_3})\bar{\Psi}(X_4,S_4)\right\rangle=\frac{1}{X_{12}^{\Delta+1/2}X_{34}^{\Delta+1/2}}\sum_{I}t_{I}g^I(U,V)\, ,
\end{equation} 
where $S$ is an auxiliary Grassmann-even spinor, related to its analogue in physical space $s$ by
\begin{equation}
S=\begin{pmatrix} x^\rho \gamma_\rho s \\ \gamma^0 s
\end{pmatrix}\, ,
\end{equation}
and
\begin{equation}
\Psi (X,\bar{S})\equiv \bar{S}\Psi(X)\, .
\end{equation}
The spinor transforming as a primary under conformal transformations is a combination of the components of the embedding space spinor, given by
\begin{equation}
\psi(x,\bar{s})=\Psi (X(x),\bar{S})\, , \quad X(x)=(x^\mu,1,x^2)\, .
\end{equation}
Therefore, all we need to compute the four-point function is to deduce the tensor structures $t_I$ in (\ref{4point_general}), project the embedding space coordinates and auxiliary spinors into physical space and finally take the corresponding derivatives with respect to $s_i$.

It turns out that there are $16$ structures of even parity\footnote{We dismiss terms of the form $$ \frac{\left\langle \bar{S}_1 X_3 S_2\right\rangle \left\langle \bar{S}_3 X_1 X_2 S_4\right\rangle}{\sqrt{X_{31}X_{32}X_{12}}}\, ,$$ as well as $$\left\langle \bar{S}_1 \Gamma S_2\right\rangle\left\langle \bar{S}_3 \Gamma S_4\right\rangle\, ,$$ which would make sense in even dimensions because $\Gamma$ is the chirality operator.} which are homogeneous of degree $0$ and scalar in embedding space. However, the fact that the correlator must be invariant under the exchange $\left\{1,2\right\}\leftrightarrow \left\{3,4\right\}$ leads to $4$ constraints, leaving $12$ independent structures,
\begin{eqnarray}
t_1&=&\frac{\left\langle\bar{S}_1 S_2\right\rangle \left\langle \bar{S}_3 X_1 X_2 S_4\right\rangle}{X_1\cdot X_2}+\frac{\left\langle\bar{S}_1 X_3 X_4 S_2\right\rangle \left\langle \bar{S}_3 S_4\right\rangle}{X_3\cdot X_4}\\
t_2&=&\frac{\left\langle \bar{S}_1 X_3 S_2\right\rangle \left\langle \bar{S}_3 X_1 S_4\right\rangle}{X_3\cdot X_1}\\
t_3&=&\frac{\left\langle\bar{S}_1 X_3 S_2\right\rangle \left\langle \bar{S}_3 X_2 S_4\right\rangle}{X_3\cdot X_2}+\frac{\left\langle\bar{S}_1 X_4 S_2\right\rangle \left\langle \bar{S}_3 X_1 S_4\right\rangle}{X_1\cdot X_4}\\
t_4&=&\frac{\left\langle \bar{S}_1 X_4 S_2\right\rangle \left\langle \bar{S}_3 X_2 S_4\right\rangle}{X_4\cdot X_2}\\
t_5&=& \left\langle \bar{S}_1 S_2\right\rangle \left\langle \bar{S}_3 S_4\right\rangle \\
t_6&=& \frac{\left\langle\bar{S}_1 X_3 X_4 S_2\right\rangle \left\langle \bar{S}_3  X_1 X_2 S_4\right\rangle}{(X_3\cdot X_4)(X_1\cdot X_2)}\\
t_{6+i}&=& t_i(\left\{1\right\}\leftrightarrow \left\{3\right\})\, ,\quad i=1, \ldots , 6
\end{eqnarray}
Hence, the tensors associated to the components $\left\langle \psi_a (x_1)\bar{\psi}_b(x_2)\psi_c (x_3)\bar{\psi}_d(x_4)\right\rangle$ are
\begin{eqnarray}
\left(t_1\right)_{abcd} &=&\frac{1}{x_{12}^2}\left[ (x_{21})_\mu (x_{13})_\nu (x_{21})_\rho (x_{42})_\sigma (\gamma^\mu)_{ab}(\gamma^\nu)_{ce}(\gamma^\rho)_{ef}(\gamma^\sigma)_{fd}\right]\nonumber\\
&+&\frac{1}{x_{34}^2}\left[ (x_{31})_\mu (x_{43})_\nu (x_{24})_\rho (x_{43})_\sigma (\gamma^\mu)_{ae}(\gamma^\nu)_{ef}(\gamma^\rho)_{fb}(\gamma^\sigma)_{cd}\right]\\
\left(t_2\right)_{abcd}&=&\frac{1}{x_{31}^2}\left[(x_{31})_\mu (x_{23})_\nu (x_{13})_\rho (x_{41})_\sigma(\gamma^\mu)_{a e}(\gamma^\nu)_{e b}(\gamma^\rho)_{cf}(\gamma^\sigma)_{fd}\right]\\
\left(t_3\right)_{abcd} &=&\frac{1}{x_{32}^2}\left[(x_{31})_\mu (x_{23})_\nu (x_{23})_\rho (x_{42})_\sigma(\gamma^\mu)_{a e}(\gamma^\nu)_{e b}(\gamma^\rho)_{cf}(\gamma^\sigma)_{fd}\right]\nonumber\\
&+&\frac{1}{x_{41}^2}\left[(x_{41})_\mu (x_{24})_\nu (x_{13})_\rho (x_{41})_\sigma(\gamma^\mu)_{a e}(\gamma^\nu)_{e b}(\gamma^\rho)_{cf}(\gamma^\sigma)_{fd}\right]\\
\left(t_4\right)_{abcd} &=&\frac{1}{x_{42}^2}\left[(x_{41})_\mu (x_{24})_\nu (x_{23})_\rho (x_{42})_\sigma(\gamma^\mu)_{a e}(\gamma^\nu)_{e b}(\gamma^\rho)_{cf}(\gamma^\sigma)_{fd}\right]\\
\left(t_5\right)_{abcd} &=&(x_{21})_\mu (x_{43})_\nu (\gamma^\mu)_{ab} (\gamma^\nu)_{cd}\\
\left(t_6\right)_{abcd} &=&\frac{1}{x_{34}^2 x_{12}^2}\left[ (x_{31})_\mu (x_{43})_\nu (x_{24})_\rho (x_{13})_\sigma (x_{21})_\theta (x_{42})_\zeta \right.\nonumber\\
& \times &\left.(\gamma^\mu)_{ae}(\gamma^\nu)_{ef}(\gamma^\rho)_{fb} (\gamma^\sigma)_{cg} (\gamma^\theta)_{gh} (\gamma^\zeta)_{hd}\right]
\end{eqnarray}
Substituting this expansion for the four-point correlator in terms of arbitrary functions of the cross ratios in (\ref{term1}), and with the help of a computer program, we calculated all the contractions involved in the four-partite information and found that it is a linear combination of too many terms to be written here, even in the specific configuration in which all regions lie in a plane. When we further arrange the regions, for example, in the vertices of a square, we schematically get
\begin{equation}
I_4 \approx c_1 \frac{R^{8\Delta}}{r^{8\Delta}} \times \left( \sum_{I,J}^{12} \#(I,J) g_I(u,v) g_J(u,v) \right)+ c_{\text{free}}\text{`two point functions'}\, .
\end{equation}

\section{$N$-partite information for $N\geq 5$}\label{sec:N_parite}

The general structure of the $N$-partite information in the long-distance approximation (\ref{IN_final}) suggests that, when the regions are spheres of radius $R$ separated a distance of order $r$, the leading behaviour should be 
\begin{equation}
I_N = \# \left(\frac{R}{r}\right)^{2 N \Delta}+ \ldots\, ,
\end{equation}
with $\Delta$ the lowest scaling dimension amongst the CFT primaries, which we assumed to be a fermion.  However, in section \ref{sec:tripartite} we showed that the coefficient $\#$ above exactly vanishes for $N=3$. Thus, given that the tripartite information scales with a power of $R/r$ grater than $2 N \Delta$, one may wonder whether this is just a particular case of a more general feature. The obvious guess is that the coefficient may vanish whenever $N$ is odd. We will first address this question in the most general scenario of fermionic CFTs, and we will later discuss what further conclusions can be drawn if the fermion is free.
 
The most insightful approach to study the general case is to organize the contributions in graphs and focus on the product of $N$ $b_{ij}$ coefficients associated to each of these diagrams. For example, a closer inspection of (\ref{I3_intermediate_step}), stemming from the triangle graph in Fig.\,\ref{N3diagram_I3}, indicates that
\begin{equation}
\sum_{i\neq j \neq k}b_{i j}b_{j k}b_{k i}=0\, ,
\end{equation}
namely, we can conclude that the tripartite information leading term vanishes  without the need to analyze the spinorial structure of the correlators. 
The above identity can be made manifest provided we introduce a skew-symmetric $N \times N$ matrix with components $[b]_{i j}=b_{i j}$. Written in terms of this matrix, we have
\begin{equation}
\text{Tr}\[ b^3\]=0\, ,
\end{equation}
which is simply a consequence of the anti-symmetry of $b^3$. 

\begin{figure}[t!]
\centering
\begin{tikzpicture}
\begin{scope}[ thick,decoration={
    markings,
    mark=at position 0.5 with {\arrow{>}}}
    ] 
\draw [postaction={decorate}] (0,-2) --(1,0);
 \draw [postaction={decorate}] (1,0)-- (-1,0); 
 \draw [postaction={decorate}] (-1,0) --(0,-2);
\draw [postaction={decorate}] (-1,0) --(0,-2);
\draw [postaction={decorate}](-1,0) .. controls (-0.5,0.5) and (0.5,0.5).. (1,0);
\end{scope}
\begin{scope}[ thick,decoration={
    markings,
    mark=at position 0.5 with {\arrow{<}}}
    ] 
\draw [postaction={decorate}](-1,0) .. controls (-0.5,-0.5) and (0.5,-0.5).. (1,0);
  \end{scope}
\draw [fill] (0,-2) circle[radius=0.06]; 
\draw [fill] (1,0) circle[radius=0.06];
\draw [fill] (-1,0) circle[radius=0.06]; 
\node [below] at (0,-2) {$k$};
\node [right] at (1,0) {$j$};
\node [left] at (-1,0.1) {$i$};
\end{tikzpicture}
\begin{tikzpicture}
\begin{scope}[ thick,decoration={
    markings,
    mark=at position 0.5 with {\arrow{>}}}
    ] 
\draw [postaction={decorate}] (0,-2) --(1,0);
 \draw [postaction={decorate}] (1,0)-- (-1,0); 
 \draw [postaction={decorate}] (-1,0) --(0,-2);
\draw [postaction={decorate}](0,-2) .. controls (0.5,-2.5) and (1.5,-2.5).. (2,-2);
\end{scope}
\begin{scope}[ thick,decoration={
    markings,
    mark=at position 0.5 with {\arrow{<}}}
    ] 
\draw [postaction={decorate}](0,-2) .. controls (0.5,-1.5) and (1.5,-1.5).. (2,-2);
  \end{scope}
\draw [fill] (0,-2) circle[radius=0.06]; 
\draw [fill] (1,0) circle[radius=0.06];
\draw [fill] (-1,0) circle[radius=0.06]; 
\draw [fill] (2,-2) circle[radius=0.06]; 
\node [below] at (0,-2) {$k$};
\node [right] at (1,0) {$j$};
\node [left] at (-1,0.1) {$i$};
\node [above] at (2,-2) {$\ell$};
\end{tikzpicture}
\begin{tikzpicture}
\begin{scope}[thick,decoration={
    markings,
    mark=at position 0.5 with {\arrow{>}}}
    ] 
\draw [postaction={decorate}] (0,0) --(0,1.4);
 \draw [postaction={decorate}] (0,1.4)-- (1,2.3); 
 \draw [postaction={decorate}] (1,2.3) --(2,1.4);
 \draw [postaction={decorate}] (2,0) --(0,0);
  \draw [postaction={decorate}] (2,1.4) --(2,0);
  \end{scope}
\draw [fill] (0,0) circle[radius=0.06]; 
\draw [fill] (0,1.4) circle[radius=0.06];
\draw [fill] (2,1.4) circle[radius=0.06];
\draw [fill] (2,0) circle[radius=0.06];
\draw [fill] (1,2.3) circle[radius=0.06]; 

\node [below] at (0,0) {$m$};
\node [above] at (0,1.4) {$i$};
\node [above] at (2,1.4) {$k$};
\node [below] at (2,0) {$\ell$};
\node [above] at (1,2.3) {$j$};
\end{tikzpicture}
\captionsetup{width=0.9\textwidth}
\caption{Connected graphs contributing to five-partite information. Since we are exclusively interested in the coefficient defined by the contraction of 5 $b_{i j}$ which is associated to each graph, we do not pay attention to the assignation of letters (regions) to paths.} 
\label{I_5}
\end{figure}
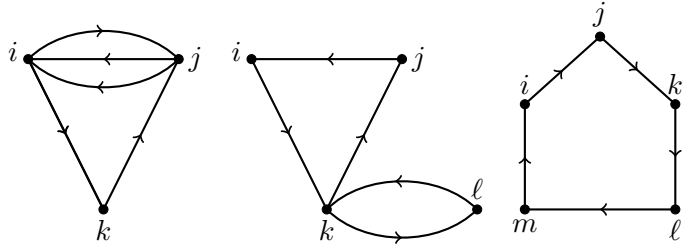

Following the same strategy, it is straightforward to show that
\begin{equation}\label{I5v}
    I_5 =0+ \dots \, , \quad  (J=1/2)
\end{equation}
for general fermionic CFTs. In order to prove this, we take into account the connected graphs, sketched in Fig.\,\ref{I_5}, as the disconnected ones must vanish due to the clustering principle, as explained in section \ref{sec:fourpartite}. The coefficients associated to the graph on the left yield
\begin{equation}
\text{Tr}\[ b^2 \cdot b^{\circ 3} \]=0\, ,
\end{equation}
where we used the element-wise product $\left[A \circ B \right]_{i j} \equiv A_{i j} B_{i j} $ to define the Hadamard power $ \left[b^{ \circ 3}\right]_{i j}= \left(b_{i j}\right)^3$. Once again, the trace is zero due to the fact that $b^{\circ 3}$ is skew symmetric, while $b^2$ is symmetric. Analogously, the contribution related to the diagram in the middle is proportional to 
\begin{equation}
\text{Tr}\left[ b^3 \circ b^2\right] =0\, ,   
\end{equation}
a consequence of the antisymmetry of $b^3 \circ b^2$. Finally, the right graph leads to
\begin{equation}
\text{Tr}\left[ b^5\right]=0\, ,
\end{equation}
meaning that (\ref{I5v}) holds, and the five-partite information goes to zero faster than expected when $r\gg R$, just like the tripartite information does. 

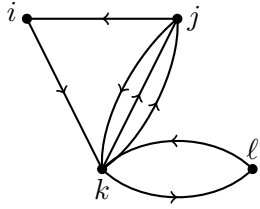
\begin{figure}[t!]
\centering
\begin{tikzpicture}
\begin{scope}[ thick,decoration={
    markings,
    mark=at position 0.5 with {\arrow{>}}}
    ] 
\draw [postaction={decorate}] (0,-2) --(1,0);
\draw [postaction={decorate}](0,-2) .. controls (0.55,-1.65) and (1.05,-0.65) .. (1,0);
 \draw [postaction={decorate}] (1,0)-- (-1,0); 
 \draw [postaction={decorate}] (-1,0) --(0,-2);
\draw [postaction={decorate}](0,-2) .. controls (0.5,-2.5) and (1.5,-2.5).. (2,-2);
\end{scope}
\begin{scope}[ thick,decoration={
    markings,
    mark=at position 0.5 with {\arrow{<}}}
    ] 
\draw [postaction={decorate}](0,-2) .. controls (0.5,-1.5) and (1.5,-1.5).. (2,-2);
  \end{scope}
  \begin{scope}[ thick,decoration={
    markings,
    mark=at position 0.5 with {\arrow{<}}}
    ] 
  \draw [postaction={decorate}](0,-2) .. controls (-0.05,-1.35) and (0.45,-0.45) .. (1,0);
  \end{scope}
\draw [fill] (0,-2) circle[radius=0.06]; 
\draw [fill] (1,0) circle[radius=0.06];
\draw [fill] (-1,0) circle[radius=0.06]; 
\draw [fill] (2,-2) circle[radius=0.06]; 
\node [below] at (0,-2) {$k$};
\node [right] at (1,0) {$j$};
\node [left] at (-1,0.1) {$i$};
\node [above] at (2,-2) {$\ell$};
\end{tikzpicture}
\captionsetup{width=0.9\textwidth}
\caption{Connected graph giving rise to a non vanishing contribution for the seven-partite information.} 
\label{I_7}
\end{figure}

Moving to $N=7$, a close inspection reveals that there is one out of 13 connected diagrams whose contribution does not vanish in general. The graph is shown in Fig.\,\ref{I_7}, and leads to the coefficient
\begin{equation}
\text{Tr}\left[ \left( b^2 \cdot b^{\circ 3}\right) \circ b^2\right]\neq 0\, .
\end{equation}
Therefore, for general CFTs the leading term of $I_N$ does not necessarily vanish when $N$ is odd and $N\geq 7$. 

However, if we further assume that the the theory is free, then the analysis is simplified, because only the product of $N$ two-point functions contributes. For that reason, it is easy to anticipate that $I_N=0$ for odd $N$. The argument goes as follows. Since each correlator (\ref{2pointfunction}) is purely imaginary and the coefficients calculated from the contraction of $b_{ij}$ are real, the combination of traces of products of gamma matrices that stem from the different permutations of regions must cancel out. This can be seen explicitly in (\ref{I3_intermediate_step}), where the permutation of the regions $B$ and $C$ associated to the free sides of the triangle (Fig.\,\ref{N3diagram_I3}) leads to a combination of two products of gamma matrices in reverse order and with opposite sign, yielding $I_3=0$. In turn, when $N=5$ there are $4!$ terms which come from the different ways of labeling $4$ sides in the pentagon of Fig.\,\ref{I_5}. These can be grouped into $12$ pairs, each giving again a product of gamma matrices and the same product but in reverse order and with opposite sign. Since this applies to arbitrarily large odd $N$, it follows that for the free fermion $N$-partite information decays with a power of $R/r$ grater than $2 N \Delta$ when $N$ is odd.

\section{Final comments}\label{sec:final_comments}
In this article we have studied the long-distance $N$-partite information of CFTs with spin-$1/2$ fields as their lowest-dimensional primary, focusing on the case corresponding to spherical regions. The results, which are summarized in the introduction, contribute to the program of reconstructing the CFT data from entanglement measures. On the one hand, unlike the scalar case, the leading term in $I_3$ vanishes identically, so four-partite information is the simplest generalization to mutual information which provides access to the structure constants of the theory at leading order. Indeed, we show that $I_4$ can be expressed in terms of both two and four-point correlators. When the theory is free, this reduces to a simple analytical formula. 
The fact that the overall coefficient in the $I_4$ has no definite sign distinguishes the free fermion from the free scalar case. Actually, at least in $d=3$, the free-scalar $I_N$ was conjectured to be positive semi-definite for all $N$ in this regime \cite{Agon:2022efa}. Naturally, multipartite information with arbitrary odd $N$ is yet a different probe for the Lorentz representation of the lightest primary. In this regard, we proved that $I_5$ vanishes at leading order for general fermionic CFTs, and that this continues to be so for $I_N$ with $N=7, 9, \dots$ if the fermion is free. 

It would be interesting to extend our results to an arbitrary representation of the Lorentz group. Namely, it would be worth studying how $N$-partite information, at leading order in the long-distance regime, depends on the spacetime directions characterizing the geometric arrangement of regions as we modify the spin of the lowest lying primary, much in the same way as \cite{Casini:2021raa} generalized the previously known mutual-information expressions. 

Another aspect that would be interesting to explore is the subleading corrections in the long-distance approximation to mutual information. The conformal block decomposition of mutual information \cite{Long:2016vkg, Chen:2016mya, Chen:2017hbk} means that these corrections must encode the spectra of the theory as well as the OPE coefficients. A very recent article \cite{Agon:2024vif} has proposed a kernel expansion of the R\'enyi twist operator that leads to a non-perturbative result, valid at all distance scales, where the contribution of all bilocal primaries and their descendants is resummed. It would be interesting to analyse if a similar scheme could be applied in the context of spin $1/2$ primaries, and what results it would lead to. 

In the holographic context, as argued in \cite{Agon:2022efa}, the equivalence between the $N$-partite information of the boundary theory and the $N$-partite information of the dual bulk theory follows from the equivalence between correlators of boundary and bulk twist operators at long distances. Such equivalence is guaranteed by the extrapolate dictionary connecting bulk and boundary correlators, and the fact that bulk and boundary modular flows coincide for near-boundary points\footnote{Remember that we are in the regime of long separation distances, which can be achieved by having finite separations but taking the limit of zero radii for all the spheres. From the bulk perspective, we are thus dealing with near-boundary observables, which makes this match quite natural. }. The simplicity of this argument implies that the previous equality must hold for any compact region, not necessarily having to be spherical provided that they are sufficiently far apart or equivalently small. On the other hand, provided one can consistently construct a holographic CFT with a free-fermion sector as the operator of the lowest scaling dimension in the theory, our findings imply that the holographic $4$-partite information cannot have a definite sign at all distances. Of course, these statements refer to the non-geometric part of the answer due to the phase transition in the RT surface happening at sufficiently large separations\cite{Headrick:2010zt}. Likewise, the 4-partite information is always computed in the ground state of the theory.

Finally, it would be interesting to explore whether there exists an interacting theory of fermions for which 
$I_4=0$ for all geometric configurations. Based on previous work on the extensive mutual information (EMI) model, which satisfies 
$I_3
 =0$, we suspect the answer to be negative. Moreover, if one could prove that statement, such proof would provide an independent argument against the physical realizability of the EMI for more than two space-time dimensions \cite{Agon:2021zvp}. Even an argument restricted to the case of spheres would be interesting.

\section*{Acknowledgements}
We  thank Horacio Casini and Gonzalo Torroba for useful discussions.
PB was supported by a Ramón y Cajal fellowship (RYC2020-028756-I) from Spain's Ministry of Science and Innovation and by the grant PID2022-136224NB-C22, funded by MCIN/AEI/ 10.13039/501100011033/FEDER, UE. PB and GvdV were supported by a Proyecto de Consolidaci\'on Investigadora (CNS2023-143822) from Spain's Ministry of Science, Innovation and Universities. The work of CA is supported by the Netherlands Organisation for Scientific Research (NWO) under the VICI grant VI.C.202.104.
\appendix

\section{Charge conjugation invariance of the twist operator}\label{charge_conj}
The charge conjugation matrix $C$ defines the following similarity transformation for the generators of the Clifford algebra $C l_{1,d-1}(\mathbf{R})$
\begin{equation}
C^{-1}\gamma^\mu C = -(\gamma^\mu)^t\, .
\end{equation}
Moreover, the charge conjugated Dirac spinor reads
\begin{equation}
\psi^c =C \bar{\psi}^t = C (\gamma^0)^t \psi^*\, .
\end{equation}
Consequently, the bilinear $\bar{\psi}_j \gamma^\mu \psi_i$ transforms as
\begin{equation}
\begin{split}
\bar{\psi}_i^c \gamma^\mu \psi_j^c &= \psi_i^t (\gamma^0)^t C^{-1} \gamma^0 \gamma^\mu C (\gamma^0)^t \psi^*_j\\
&= \psi_i^t (\gamma^0)^t (\gamma^0)^t (\gamma^\mu)^t (\gamma^0)^t \psi^*_j\\
&=-\bar{\psi}_j\gamma^\mu \psi_i\, ,
\end{split}
\end{equation}
where in going from the second to the third line a minus sign accounts for the anti-commutation of the spinors. This shows that (\ref{twist_OPE}) is charge-conjugation invariant.

\section{Fermionic modular correlator \label{fmc}}

In this appendix we present a derivation of (\ref{numerator}) following \cite{Casini:2021raa}. We start with the following fermionic correlator ${\langle \Omega |\Sigma^{(n)}_A \bar{\psi}_\alpha^{k}(x_1) {\psi}_\beta^{l}(x_2)|\Omega\rangle}$. For simplicity, we assume $n>k>l\geq 1$ and $x_i\in A$. Intuitively, in a replica realization of this correlator, one can write 
\bea\label{corr-rep-ferms}
\frac{\langle \Omega |\Sigma^{(n)}_A \bar{\psi}_\alpha^{k}(x_1) {\psi}_\beta^{l}(x_2)|\Omega\rangle}{\langle \Omega |\Sigma^{(n)}_A|\Omega\rangle}=\frac{\Tr\{\rho_A^{n-k} \bar{\psi}_\alpha(x_1)\rho_A^{k-l}{\psi}_\beta(x_2)\rho_A^{l}\}}{\Tr \rho^n_A}\,.
\eea
Since the reduced density matrix is bounded $0\leq \rho_A\leq 1$, the operator $\rho_A^a$ for positive $a$ is likewise bounded, and thus, the right-hand side representation of the above correlator is well defined.
The right-hand side of (\ref{corr-rep-ferms}) can be expressed in terms of modularly evolved correlators, defined as 
\bea\label{mod-evol-ferm}
\psi_\beta[x,s]=\Delta^{-is}\psi_\beta (x)\Delta^{is}
\eea
where in finite dimensional systems the modular operator $\Delta$, can be expressed as $\Delta=\rho_A\otimes \rho^{-1}_{A'}$ with $A'$ the region space-like to $A$ and we assumed both $\rho_A$ and $\rho_{A'}$ are full rank operators\footnote{Notice that every time we talk about density matrices we are implicitly assuming a regulator for the QFT since in continuous QFTs density matrices or traces do not exist. Nevertheless, correlators of modularly evolved operators are well defined in continuum QFTs, such as the one appearing on the right-hand side of (\ref{trace-mod-corr}), for example.}. In the $n\to 1$ limit, we can write 
\bea\label{trace-mod-corr}
\frac{\Tr\{\rho_A^{n-k} \bar{\psi}_\alpha(x_1)\rho_A^{k-l}{\psi}_\beta(x_2)\rho_A^{l}\}}{\Tr \rho^n_A}&\underset{n\to 1}{\approx}& \langle \Omega | \bar{\psi}_\alpha(x_1) \Delta^{\frac{k-l}{n}}{\psi}_\beta(x_2)\Delta^{-\frac{k-l}{n}}|\Omega\rangle \nonumber \\
&=&\langle \Omega | \bar{\psi}_\alpha (x_1){\psi}_\beta\left[x_2,i \tau_{kl}\right]|\Omega\rangle\,,
\eea
where $\tau_{kl}=(k-l)/n$. In the first line we use the cyclicity of the trace, and in the second, definition (\ref{mod-evol-ferm}). Finiteness of (\ref{corr-rep-ferms}) implies that the above correlator is well defined on the complex strip with $0\leq \Im (s)\leq 1 $. 
For $l<k$ we can use the anti-commutativity of the field operators and write instead the correlator 
$-\langle \Omega | {\psi}_\beta\left[x_2,i \tau_{kl}\right]\bar{\psi}_\alpha (x_1)|\Omega\rangle$ which by the same reasoning is finite and well defined. Thus, the following modular evolved correlator \cite{Haag:1967sg,Haag:1992hx}
\bea
\label{G-mod-ferms}
G_{\alpha \beta}(x_1,x_2;s):=\left\{\begin{array}{ll}
\!\!-\langle \Omega | {\psi}_\beta\left[x_2,s\right]\bar{\psi}_\alpha (x_1)|\Omega\rangle\,, & {\rm for} \,\,\, -1<\Im(s)<0\\
\,\,\langle \Omega | \bar{\psi}_\alpha (x_1){\psi}_\beta\left[x_2,s\right]|\Omega\rangle  \,, &{\rm for}\quad \, \,\,\,\,0<\Im(s)<1
\end{array} \right. 
\eea
admits an analytic continuation in the complex strip $-1<\Im(s)<1$. A fundamental property of modular correlators is the KMS condition which states
\bea
\langle \Omega | \bar{\psi}_\alpha (x_1){\psi}_\beta\left[x_2,s\right]|\Omega\rangle=\langle \Omega |{\psi}_\beta\left[x_2,s+i\right] \bar{\psi}_\alpha (x_1)|\Omega\rangle\,.
\eea
This together with the previous definition implies 
\bea
G_{\alpha \beta}(x_1,x_2;s+i)=-G_{\alpha \beta}(x_1,x_2;s)\,.
\eea
The above property allows an extension of the modular correlator to the whole complex plane. Alternatively, we can define the modular transformation of the field operator $\psi_\beta(x)$ to satisfy 
\bea\label{anti-periodic-bc}
\psi_\beta[x,s+i]=-\psi_\beta[x,s]\,,
\eea
and extend the part of definition (\ref{G-mod-ferms}) from the range $0<\Im(s)<1$ to the whole complex plane. Equation (\ref{numerator}) follows this extension and the explicit use of (\ref{anti-periodic-bc}).

\bibliographystyle{JHEP-2}
\bibliography{N-partite-info-v4}

\end{document}